\documentclass{caosp}

\usepackage{graphicx}

\articleNo{xxx} \pubyear{2006} \volume{36} \firstpage{103}
\received{December 12, 2005} \accepted{February 27, 2006}
\begin{document}

%\htitle{The Taurid complex meteor showers and asteroids}
\hauthor{V.\,Porub\v{c}an, L.\,Korno\v{s} and I.P.\,Williams}
\title{The Taurid complex meteor showers and asteroids}
\author{V.\,Porub\v{c}an \inst{1}, L.\,Korno\v{s} \inst{1}
\and I.P.\,Williams \inst{2}}
\institute{Astronomical Institute, Faculty Mathematics, Physics
and Informatics, Comenius University, 828 48 Bratislava, The
Slovak Republic \and Astronomy Unit, Queen Marry, University of London,
Mile End Road, London E1 4NS, UK}
\date{September 20, 2004}
\maketitle

\begin{abstract}
The structure of the Taurid meteor complex based on photographic orbits
available in the IAU Meteor database is studied. We have searched for potential
sub-streams or filaments to be associated with the complex utilizing the
Southworth-Hawkins D-criterion. Applying a strict limiting value for $D=$ 0.10,
fifteen sub-streams or filaments, consisting of more than three members, could
be separated out from the general complex. To confirm their mutual consistence
as filaments, rather than fortuitous clumping at the present time, the  orbital
evolution over 5000 years of each member is studied. Utilizing the D-criterion
we also searched for NEOs that might be associated with the streams and
filaments of the complex and investigated the orbital evolution of potential
members. Possible associations between 7 Taurid filaments and 9 NEOs were
found. The most probable are for S Psc(b) -- 2003\,QC10, N Tau(a) --
2004\,TG10, $o$ Ori -- 2003\,UL3 and N Tau(b) -- 2002\,XM35. Some of the
potential parent objects could be either dormant comets or larger boulders
moving within the complex. Three of the most populated filaments of the complex
may have originated from 2P/Encke. 
\end{abstract}

\section{Introduction}

The Taurids are a meteor shower that originates from a stream with very low
inclination so that the Earth is almost moving within the complex for a large
fraction of its orbit and that meteors are seen on Earth near to both the
ascending and descending nodes. This means that the Taurid complex encounters
the Earth two times with the duration of each resulting shower being very long.
Before perihelion passage it generates the night-time showers called the
Northern and Southern Taurids (September 15 -- December 1, Cook, 1973) and in
postperihelion  as the daytime showers recognized by radio observations - Zeta
Perseids and Beta Taurids (May 20 -- July 6, Sekanina, 1973). In the beginning
of the 20th century, Denning (1928) recognized the complex nature of the
stream, identifying thirteen active radiants situated in Aries and Taurus.

Porub\v{c}an and \v{S}tohl (1987) analysed all photographic data on the Taurids
available  at that date and found a very long activity period of the stream
extending in the solar longitude for almost 120 degrees.

It is generally accepted (Olsson-Steel, 1988; Babadzhanov {\it et al.}, 1990;
\v{S}tohl, Porub\v{c}an, 1990; Steel {\it et al.}, 1991; Babadzhanov, 2001,
etc.) that the stream is, in fact, a complex of several small meteor streams.
Some are genetically associated with comet P/Encke and several to Apollo
asteroids. Whether the Apollo asteroids are associated with the comet is a more
profound question that we will not discuss here.

In our analysis we have searched in the current version of the IAU Meteor Data
Center catalogue of photographic orbits (Lindblad {\it et al.}, 2005 -
available at {\it www.astro.sk/\~{}ne/IAUMDC/Ph2003/database.html}) for
meteoroid orbits belonging to the complex. We have also searched the available
data on NEO's for potential parents.

\section{The Taurid stream filaments}

The Taurids are rich in  bright meteors thus the activity and structure of the
stream is best known from photographic observations which  provide also the
best determined orbits. In his analysis of photographic Taurids, Whipple (1940)
used a sample of 14 orbits and suggested a possible relationship between the
Taurids and comet Encke. At present, the updated  IAU MDC catalogue (Lindblad
{\it et al.}, 2005) lists 4581 orbits compiled from 35 different catalogues in
which 240 members of the Taurid complex could be identified (Porub\v{c}an,
Korno\v{s}, 2002).

On the other hand, there is a steady increase in the number of the NEOs that
have been discovered, amounting to over three thousand at present.
These provide an additional source of potential co-parents of
the complex or of individual filaments.

Our approach was in two steps. In the first step a computerized stream-search
based on an iteration procedure and Southworth-Hawkins D criterion was applied
to the sample of photographic orbits covering the period of activity of the
complex, approx. September -- January.

$D$ (Southworth, Hawkins 1963) is given by

$$D_{AB}^2= (e_A-e_B)^2+(q_A-q_B)^2 + [2\sin{(I_{AB}/2)}]^2$$
$$+(e_A+e_B)^2[\sin{(\Pi_{AB}/2)}]^2.$$

Here $e_A$, $e_B$, $q_A$ and $q_B$ are the eccentricities and perihelion
distance of orbits $A$ and $B$, $I_{AB}$ is the angle between the orbital
planes and ${\Pi_{AB}}$ is the difference between the longitude of perihelion,
measured from the intersection point of the orbital planes (not from either
node). Both $I$ and $\Pi$ can be expressed in terms of the three angular
orbital elements, $i$, $\omega$ and $\Omega$, but we do not give these
expressions here.

To get the cores or central parts of the streams or sub-streams a strict
limiting value of $D\le$ 0.1 was applied. Then only filaments moving in the
orbits related to the Taurids considering the geocentric velocity
($\pm$5\,km\,s$^{-1}$ with respect to the mean geocentric velocity of the
stream) and position of the radiant ($\pm$10$\degr$ with respect to the Taurid
radiant ephemeris allowed for the radiant daily motion) were selected. In this
way 23 filaments consisting of 2 to 56 members were identified.

This procedure identified 210 meteors belonging to the complex as a whole, 84
to the Northern branch and 126 to the Southern branch. Fig.\,1 shows the
radiants of meteors of individual filaments of the Northern and Southern branch
(depicted by various symbols -- upper plot). One possible way of verifying
whether  individual filaments can be related to the Taurid complex is to reduce
the radiants to a common solar longitude, e.g. corresponding to the maximum of
the Taurid stream activity by allowing for their radiant daily motion. For
reduction, the Taurid radiant daily motion ephemeris derived by Porub\v{c}an
and Korno\v{s} (2002) was applied. The radiants  reduced to the solar longitude
of 220$\degr$ (lower plot), show the size of the radiant area for the Northern
and Southern branch. The radiant areas of both branches are well defined,
compact and form a common area of the size of about 25$\degr\times$15$\degr$.
The most outlying filament in January (filament no. 23 at right ascension
140$\degr$ -- 150$\degr$, upper plot of Fig.\,1) is close to 7\,:\,2
mean-motion resonance with Jupiter and may not belong to the Taurid complex.
%Fig.1
\begin {figure}
\centerline{\includegraphics[width=4.5cm,angle=-90]{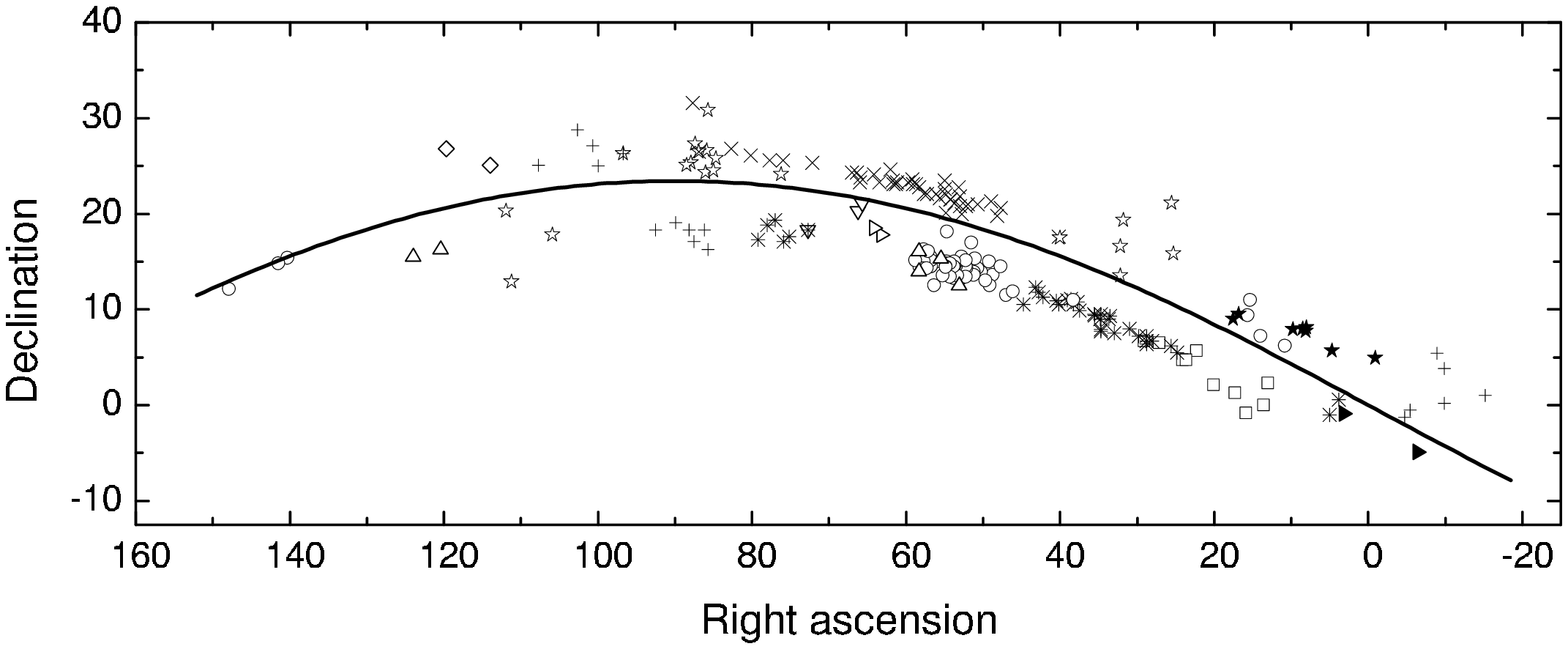}}
\vspace{0.7cm}
\centerline{\includegraphics[width=4.5cm,angle=-90]{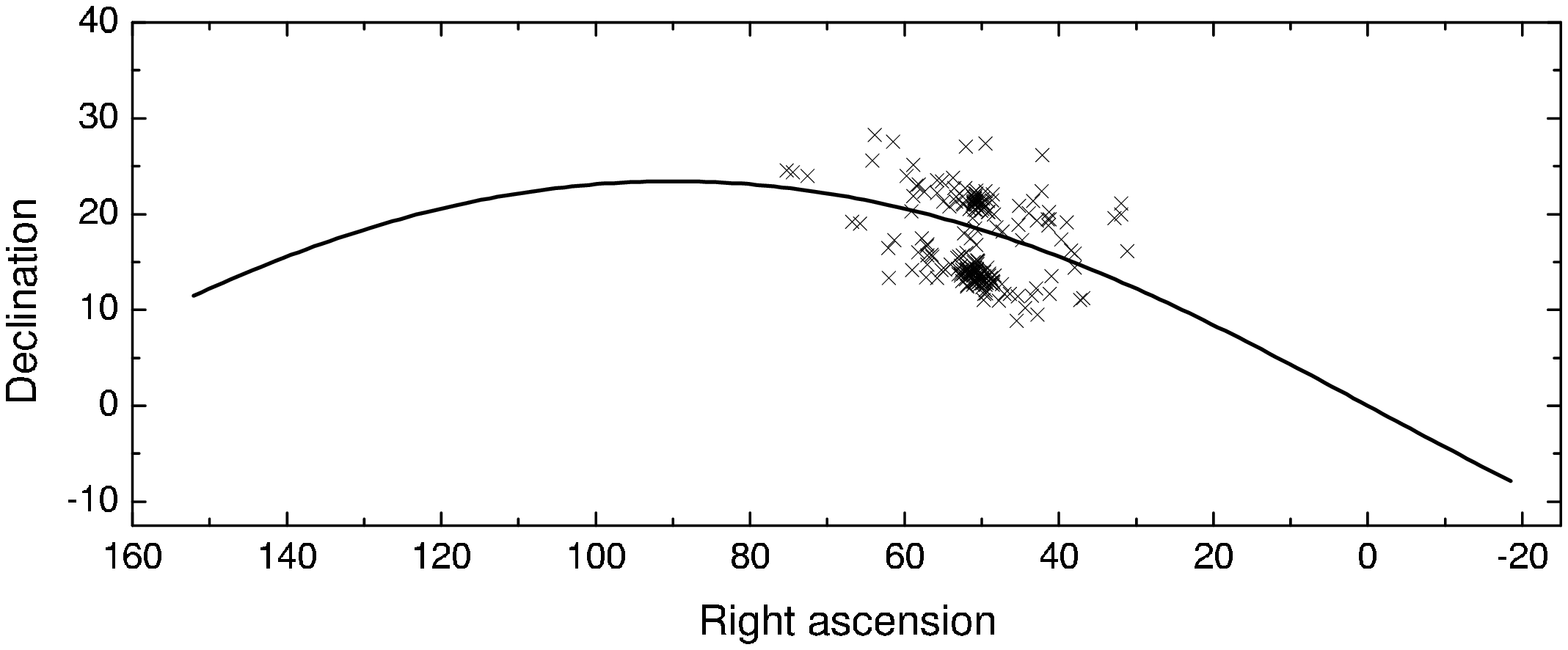}}
\caption{The
radiant positions within the Taurid meteor complex (upper plot) and the
radiants reduced to the common solar longitude of 220$\degr$ (lower plot).
Meteors of individual filaments in the upper plot are depicted by different
symbols.}
\end{figure}

   Due to a very long period over which the shower is active, about four months
(at least 120$\degr$ in the solar longitude) September -- December,  the Taurid
radiant passes through several constellations. In consequence amongst the
meteors identified by the above procedure as belonging to the Taurids are also
meteors that have been designated by the authors of other catalogues as members
of other minor streams.

   By comparing the mean orbits of the N Piscids, the S Arietids and both
branches of the Chi Orionids with the orbit of the N and S Taurids it is
apparent that these streams can be regarded as extended members of the TC
(Porub\v{c}an and \v{S}tohl, 1987).

In the later work, we have considered only filaments that contain four members
at least. This gives 15 filaments and their mean orbits, radiants, geocentric
velocities and corresponding Tisserand invariant values (with respect to the
Jupiter's orbit) are given in Tab.\,1.

%TABLE 1:
\begin{table}[h]
\small
\begin{center}
\caption{The mean orbital elements, radiants and geocentric velocities of the
Taurid complex filaments. $T$ is the Tisserand invariant value.}
\begin{tabular}{lccccccccccccc}
\hline\hline
filament &Q& q & a & e & i &$\omega$ &$\Omega$ & $\pi$ & n & R.A.& Dec.
&$V_{g}$ & T \\
    &$(AU)$&$(AU)$&$(AU)$&  &$(\degr)$&$(\degr)$&$(\degr)$&$(\degr)$& &$(\degr)$
&$(\degr)$&(km\,s$^{-1}$) & \\
\hline
%                Q    q     a      e     i   arg    node    pi    n   R.A.   Dec   Vg   T
01 N Psc(a)   &3.08&.410& 1.743& .766& 6.7& 291.8& 160.3&  92.1& 4& 349.0&  2.6& 25.06&3.73 \\
02 S Psc(a)   &3.99&.309& 2.148& .855& 2.6& 120.3& 346.6& 106.8& 4&   1.4& -1.6& 29.58&3.09 \\
03 N Psc(b)   &3.93&.310& 2.115& .857& 4.6& 300.1& 177.0& 117.1& 8&   9.1&  7.6& 29.50&3.12 \\
04 S Psc(b)   &2.59&.338& 1.465& .770& 6.1& 121.8& 359.5& 121.3& 5&  16.0&  1.0& 25.77&4.23 \\
05 S Psc(c)   &3.54&.318& 1.925& .837& 5.7& 120.1&  10.9& 131.0& 5&  25.2&  5.7& 28.73&3.37 \\
06 $\beta$ Ari&3.72&.375& 2.048& .817& 4.8& 292.7& 203.8& 136.5& 6&  31.2& 17.3& 27.33&3.26 \\
07 S Psc(d)   &3.36&.296& 1.826& .839& 6.1& 123.0&  19.8& 142.7&28&  35.0&  9.0& 29.12&3.49 \\
08 S Tau(b)   &2.85&.225& 1.539& .854& 8.2& 132.9&  36.3& 169.2& 4&  56.3& 14.5& 30.47&3.94 \\
09 S Tau(a)   &4.01&.365& 2.187& .833& 5.3& 113.1&  43.3& 156.4&56&  53.6& 14.4& 28.16&3.09 \\
10 N Tau(a)   &3.99&.358& 2.174& .835& 2.7& 293.7& 228.6& 162.4&38&  57.6& 22.4& 28.30&3.10 \\
11 $o$ Ori   &4.34&.429& 2.385& .820& 5.0& 104.8&  69.8& 174.6& 5&  77.1& 18.0& 26.84&2.96 \\
12 N Tau(b)   &4.37&.380& 2.372& .841& 2.9& 290.1& 255.7& 185.8& 9&  85.3& 26.0& 28.35&2.92 \\
13 N $\chi$ Ori&3.81&.476& 2.143& .779& 3.3& 280.4& 256.8& 177.2& 7&  81.6& 26.9& 24.73&3.23 \\
14 S $\chi$ Ori&4.91&.438& 2.673& .836& 5.6& 102.7&  81.4& 184.2& 6&  88.3& 17.9& 27.22&2.73 \\
15 $\epsilon$ Gem &3.74&.400& 2.071& .808& 3.7& 289.2& 270.6& 199.8& 5& 101.6& 26.5& 27.05&3.26 \\
\hline
\hline
\end{tabular}
\end{center}
\end{table}

\section{NEOs in the Taurid meteor complex}

It is difficult to find a simple explanation for the long duration and large
dispersion of the orbits within the TC and of its origin. Whipple (1940)
suggested that in the distant past a giant comet disintegrated gently into a
number of smaller comets, one of which was P/Encke while another of these
disintegrated into the Taurid stream. Whipple and Hamid (1952) calculated the
secular perturbations for 9 photographic Taurids and found that some of them
had to be ejected from the comet 4700 years ago and others probably from a
hypothetical companion about 1400 years ago. At present, as well as P/Encke,
several  Apollo asteroids are considered possible progenitors (Olsson-Steel,
1988; Babadzhanov {\it et al.}, 1990; \v{S}tohl and Porub\v{c}an, 1990; Asher
{\it et al.}, 1993; Steel and Asher, 1996; Babadzhanov, 2001, and others).

   Typical asteroids by their nature and structure differ from comets and
so the mechanisms for the formation of meteoroid streams from asteroids should
be different. The most basic scenarios for meteoroid ejection are collisions
(Williams, 1993), disruption of an asteroid through fast rotation, or possible
ejections due to thermal tension at close perihelion passages. However, these
mechanisms are relatively rare events and are unlikely to provide a steady
source for a regular supplying of meteoroids into a stream as is the case of
cometary streams.

   To find potential associations between Tau filaments and NEOs,
 we searched the Asteroid Orbital
Elements Database (Ted Bowell, Lowell Observatory - {\it
http://alumnus.caltech.edu/\~{}nolan/astorb.html}~) for discoveries up-to
mid-June 2005 and found 3380 objects.

   Orbital similarity was verified by comparing the mean orbits of
the filaments with the orbits of NEOs applying the Southworth-Hawkins
D-criterion. Considering that the present orbital similarities based on the
osculating elements cannot reflect potential close associations of the TC
filaments with NEOs in the past, for the analysis asteroids for which the value
of $D\leq$ 0.30 were selected and thus 91 NEOs were identified. The
theoretical meteor radiants (right ascension $\alpha$, declination $\delta$)
and the geocentric encounter velocity $V_g$ at the approach to the Earth's
orbit were then computed (Neslu\v{s}an {\it et al.}, 1998).

We have integrated the motion of all the 91 asteroids and the mean orbits of
the filaments by using the multi-step procedure of Adams-Bashforth-Moulton's
type up to 12th order, with variable step-size and positions of the perturbing
major planets were obtained from the Planetary and Lunar Ephemerides DE406
prepared by the Jet Propulsion Laboratory (Standish, 1998).

 Though meteoroid streams can be typically recognized over a time scale of
10$^{3}$  to 10$^{4}$ years (see Arter and Williams, 1997 for formulae and
calculations) we followed the orbital evolution of the bodies for 5000 years,
the time which could be count for sufficient to indicate on potential
associations between the studied objects.

In Tab.\,2 we show the resulting list of potential parents obtained by
comparing the orbital elements, radiants and velocities of filaments with the
potentially associated objects. Besides the filaments listed in Tab.\,2 an
additional close association between filament no. 6 and four asteroids was
identified. However, due to the fact that the orbit of the filament is in a close
4\,:\,1 resonance with Jupiter and thus the filament particles are on chaotic
orbits, no real $D$ value could be obtained and, therefore, this association was
not included in Tab.\,2.

Tab.\,2 lists besides the geocentric radiant and velocity also the absolute
magnitude of the asteroid $H$ and the diameter defined for the albedos in the
range 0.05 -- 0.25. The table also gives the change of $D$ value ($D_{SH}$)
over the period of integration.

%Table 2: Taurid complex filaments - NEO
\begin{table}[h]
\small
\begin{center}
\caption{Taurid complex filaments - NEO}
\begin{tabular}{cccccccc}
\hline\hline
stream & R.A.& Dec. & $V_g$ & H & Diameter & D$_{SH}$ & Period \\
       &$(\degr)$&$({\degr})$&(km/s)&(1,0)& (m)& &(yrs)  \\
\hline
%              R.A. Dec  Vg      H    Diameter      Dsh    Period
Tau 01       & 349 &  3 & 25 \\
2001 HB      & 330 & -7 & 23 & 20.50 & 210-470 & 0.15-0.25 & 5000\\
2003 SF      &  14 &  3 & 24 & 19.57 & 320-720 & 0.12-0.25 & 2000\\
\hline
Tau 02       &   1 & -2 & 29 \\
2001 QJ96    & 356 & -7 & 27 & 21.98 & 110-240 & 0.15-0.13 & 3000  \\
\hline
Tau 04       &  16 &  1 & 26 \\
1999 RK45    &  16 &  9 & 26 & 19.32 & 360-810 & 0.19-0.19 & 5000\\
2003 QC10    &  16 &  2 & 24 & 17.83 & 720-1610& 0.07-0.06 & 5000\\
\hline
Tau 09       &  54 & 14 & 28 \\
2003 WP21    &  67 & 17 & 25 & 21.43 & 140-310 & 0.20-0.15 & 5000\\
\hline
Tau 10       &  58 & 22 & 28 \\
2004 TG10    &  55 & 22 & 30 & 19.40 & 350-780 & 0.25-0.05 & 5000\\
\hline
Tau 11       &  77 & 18 & 27 \\
2003 UL3     &  67 & 22 & 26 & 17.85 & 720-1600& 0.25-0.05 & 3200\\
2003 WP21    &  67 & 17 & 25 & 21.43 & 140-310 & 0.25-0.15 & 4000\\
\hline
Tau 12       &  85 & 26 & 28 \\
2002 XM35    &  81 & 26 & 28 & 22.96 &  70-150 & 0.25-0.05 & 3300\\
\hline
\hline
\end{tabular}
\end{center}
\end{table}

%TABLE asteroids and comet 2P/Encke
\begin{table}[h]
\small
\begin{center}
\caption{The orbital elements of Taurid complex asteroids and comet 2P/Encke.
$T$ is the Tisserand invariant value.}
\begin{tabular}{lccccrrrrc}
\hline\hline object    &~  Q   &~  q   &~  a   &~ e &   i~~
&$\omega$~~&$\Omega$~~&
$\pi$~~  &~ T \\
          &~$(AU)$&~$(AU)$&~$(AU)$&
&$(\degr)$&$(\degr)$~&$(\degr)$~&$(\degr)$~&    \\
\hline
% name        Q        q        a        e        i     Peri     Node      Pi

1999 RK45 &~ 2.834 &~ 0.363 &~ 1.598 &~ 0.773 &~ 5.9 &~   4.0 &~ 120.1 &~ 124.1
&~ 3.96 \\
2001 HB   &~ 2.225 &~ 0.402 &~ 1.314 &~ 0.694 &~ 9.3 &~ 237.7 &~ 196.1 &~  73.7
&~ 4.68 \\
2001 QJ96 &~ 2.876 &~ 0.321 &~ 1.599 &~ 0.799 &~ 5.9 &~ 121.3 &~ 339.1 &~ 100.4
&~ 3.92 \\
2002 XM35 &~ 4.212 &~ 0.378 &~ 2.295 &~ 0.835 &~ 3.1 &~ 312.8 &~ 229.8 &~ 182.6
&~ 3.00 \\
2003 QC10 &~ 2.379 &~ 0.369 &~ 1.374 &~ 0.731 &~ 5.0 &~ 120.5 &~   0.3 &~ 120.8
&~ 4.49 \\
2003 SF   &~ 3.842 &~ 0.481 &~ 2.162 &~ 0.778 &~ 5.7 &~  31.8 &~  77.6 &~ 109.5
&~ 3.22 \\
2003 UL3  &~ 4.036 &~ 0.457 &~ 2.246 &~ 0.797 &~14.6 &~  13.0 &~ 153.2 &~ 166.1
&~ 3.09 \\
2003 WP21 &~ 4.115 &~ 0.489 &~ 2.302 &~ 0.788 &~ 4.3 &~ 123.7 &~  38.1 &~ 161.8
&~ 3.08 \\
2004 TG10 &~ 4.169 &~ 0.315 &~ 2.242 &~ 0.860 &~ 3.7 &~ 310.0 &~ 212.3 &~ 162.3
&~ 2.99 \\
\hline
2P/Encke  &~ 4.097 &~ 0.339 &~ 2.218 &~ 0.847 &~11.8 &~ 186.5 &~ 334.6 &~ 161.1
&~ 3.03 \\
2005 TF50 &~ 4.247 &~ 0.292 &~ 2.269 &~ 0.871 &~10.7 &~ 159.8 &~   0.8 &~ 160.5
&~ 2.93 \\
\hline
\hline
\end{tabular}
\end{center}
\end{table}

The results of integration for probable associations are depicted in
Figs.\,3-7, where the plots of the orbital elements - $a, q, e, i, \omega$ and
$\Omega$  of the objects as well as of the differences between the lines of
apsides $\Delta\pi$, the values of $D$ and their evolution in the last 5000
years are presented. For the integration, the mean orbit of the stream was
represented by 18 modeled particles distributed equidistantly along the orbit
of the stream in the mean anomaly by 20 degrees.

The mean radiants of the filaments listed in Tab.\,1 (open circles) and
positions of the theoretical meteor radiants of the associated NEOs (stars)
calculated for the osculating orbits are plotted in Fig.\,2. The orbital
elements of the NEOs are listed in Tab.\,3.

%Fig.2: ast-rad
\begin {figure}
\centerline{\includegraphics[width=4.5cm,angle=-90]{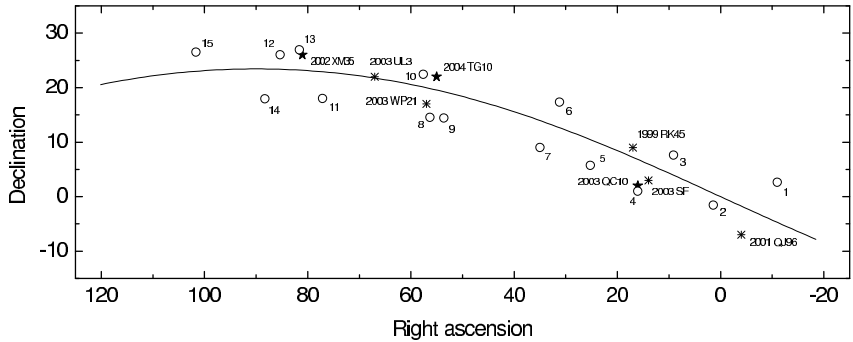}}
\caption{The mean radiants of the TC filaments (open circles) and
theoretical meteor radiants of the associated NEOs (stars).}
\end{figure}

\section{Discussion and conclusions}

Based on the similarity of the orbital evolution, the TC sub-streams
(filaments) can be arranged into four groups, which at the same time correspond
to different values of the Tisserand invariant $T$ (with respect to Jupiter's
orbit). In order of the increasing $T$ (Tab.\,1) can be arranged as: (I) - the
first group  with $T\le3.0$ is formed by filaments 11, 12 and 14; (II) - the
second and the most numerous group has $T$ in the interval of 3.09 -- 3.26 and
consists of filaments 2, 3, 6, 9, 10, 13 and 15; (III) - the third group is formed
by two filaments 5 and 7 with $T$ between 3.37 -- 3.49; (IV) - the fourth group
consisting of filaments 1, 4 and 8, has the largest $T$ between 3.73 -- 4.23.

%Fig. Tau 04
\begin {figure}
\centerline{\includegraphics[width=3.3cm,angle=-90]{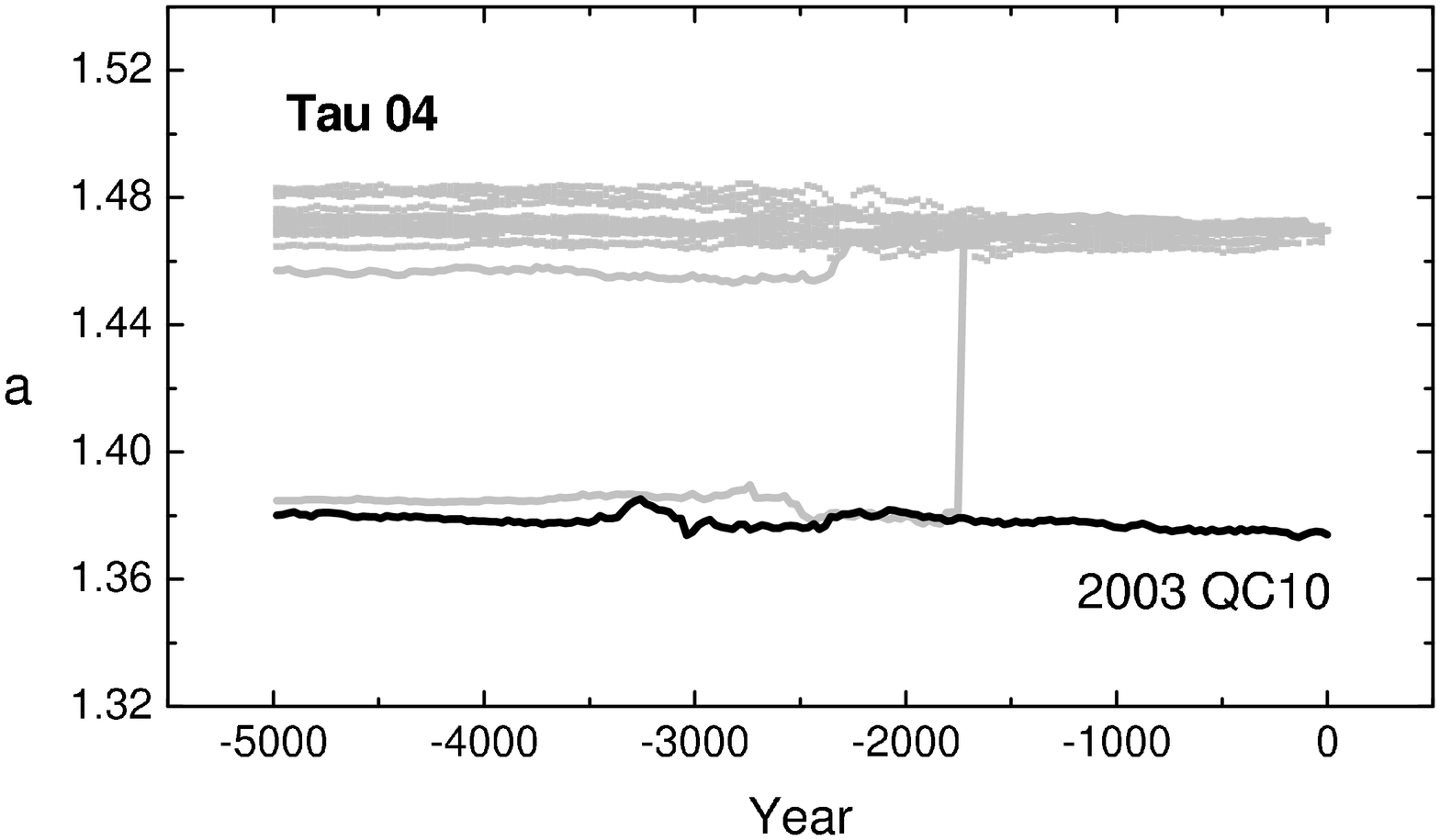}
           \hspace{0.3cm}
            \includegraphics[width=3.3cm,angle=-90]{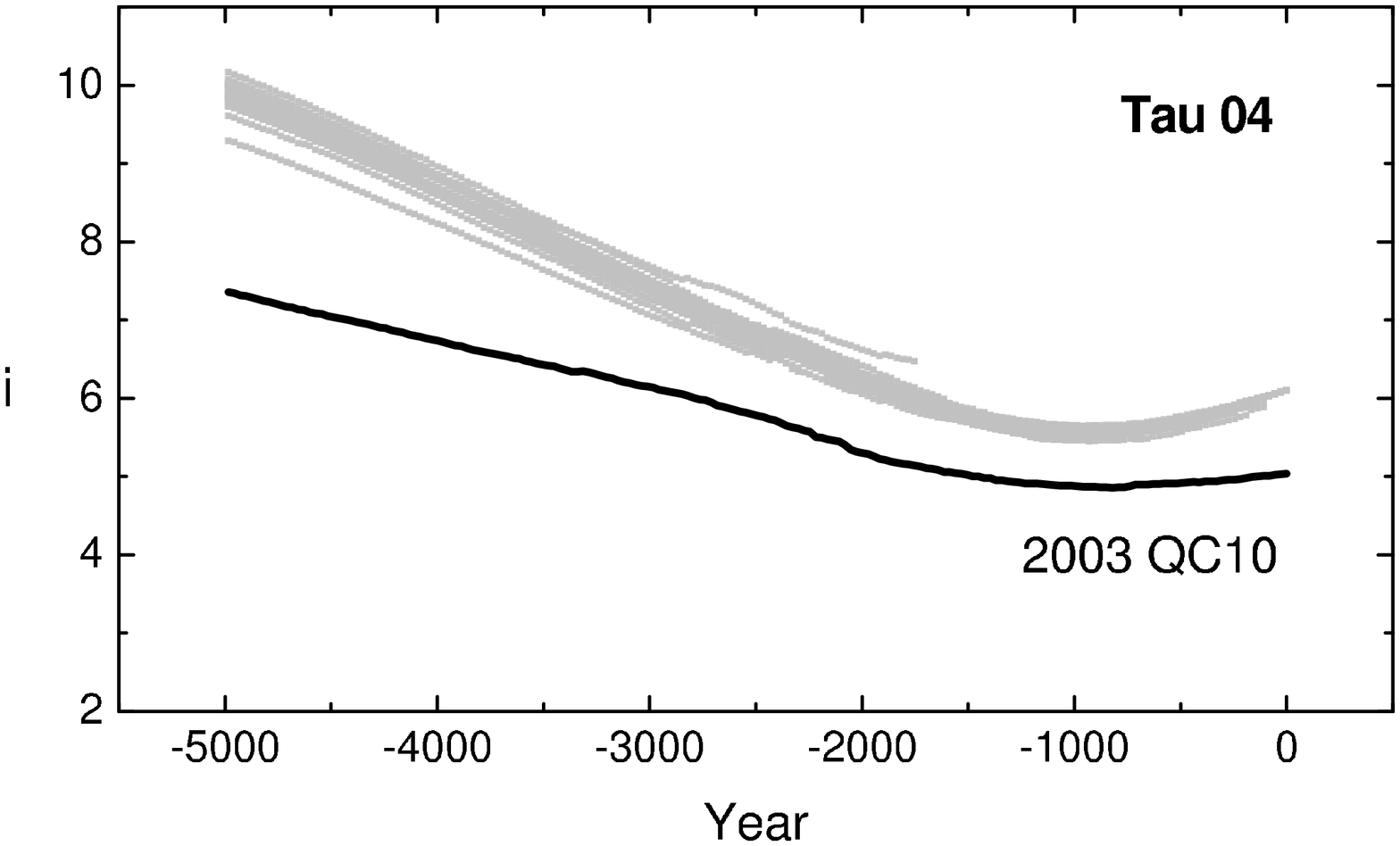}
           }
\vspace{0.7cm}
\centerline{\includegraphics[width=3.3cm,angle=-90]{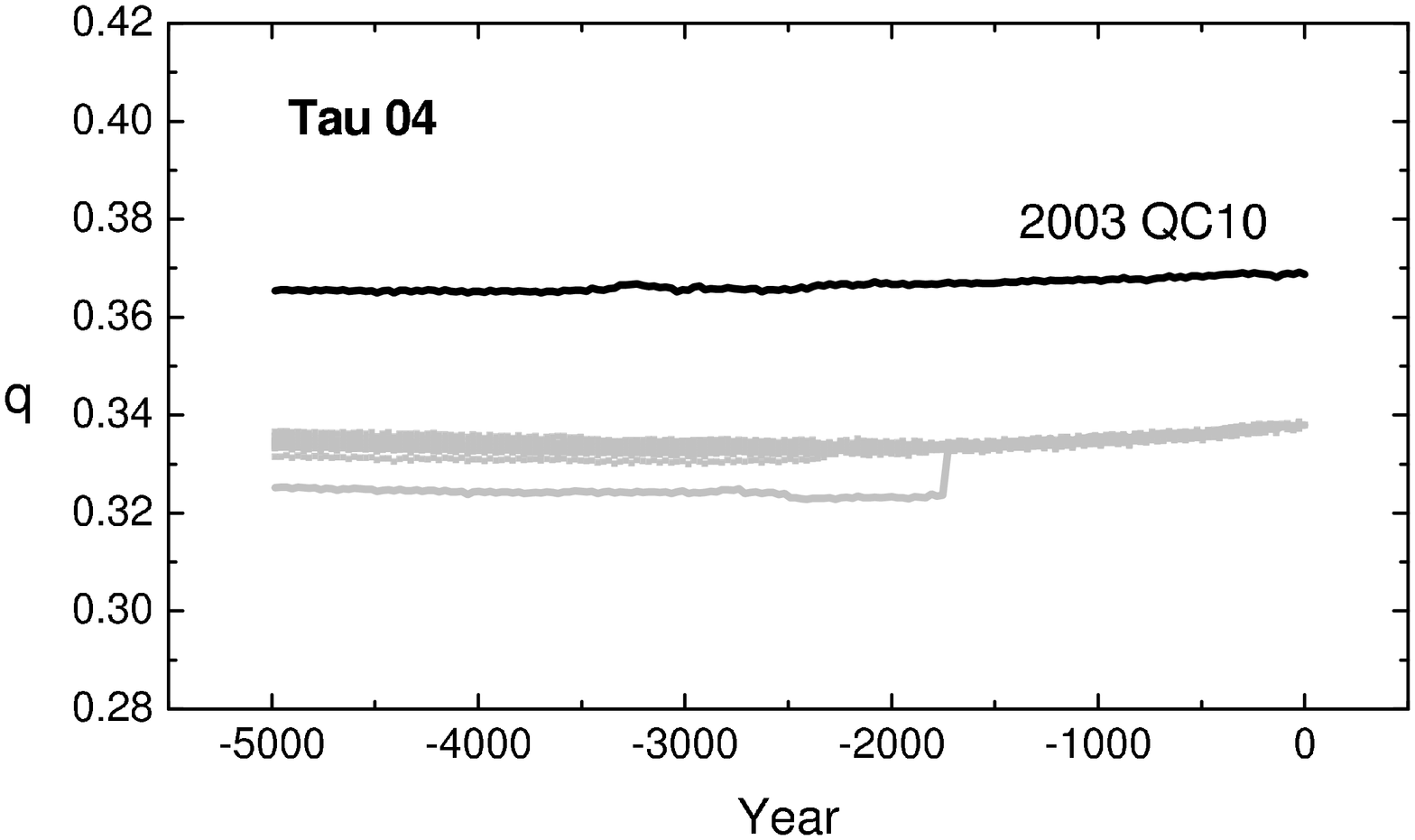}
            \hspace{0.3cm}
            \includegraphics[width=3.3cm,angle=-90]{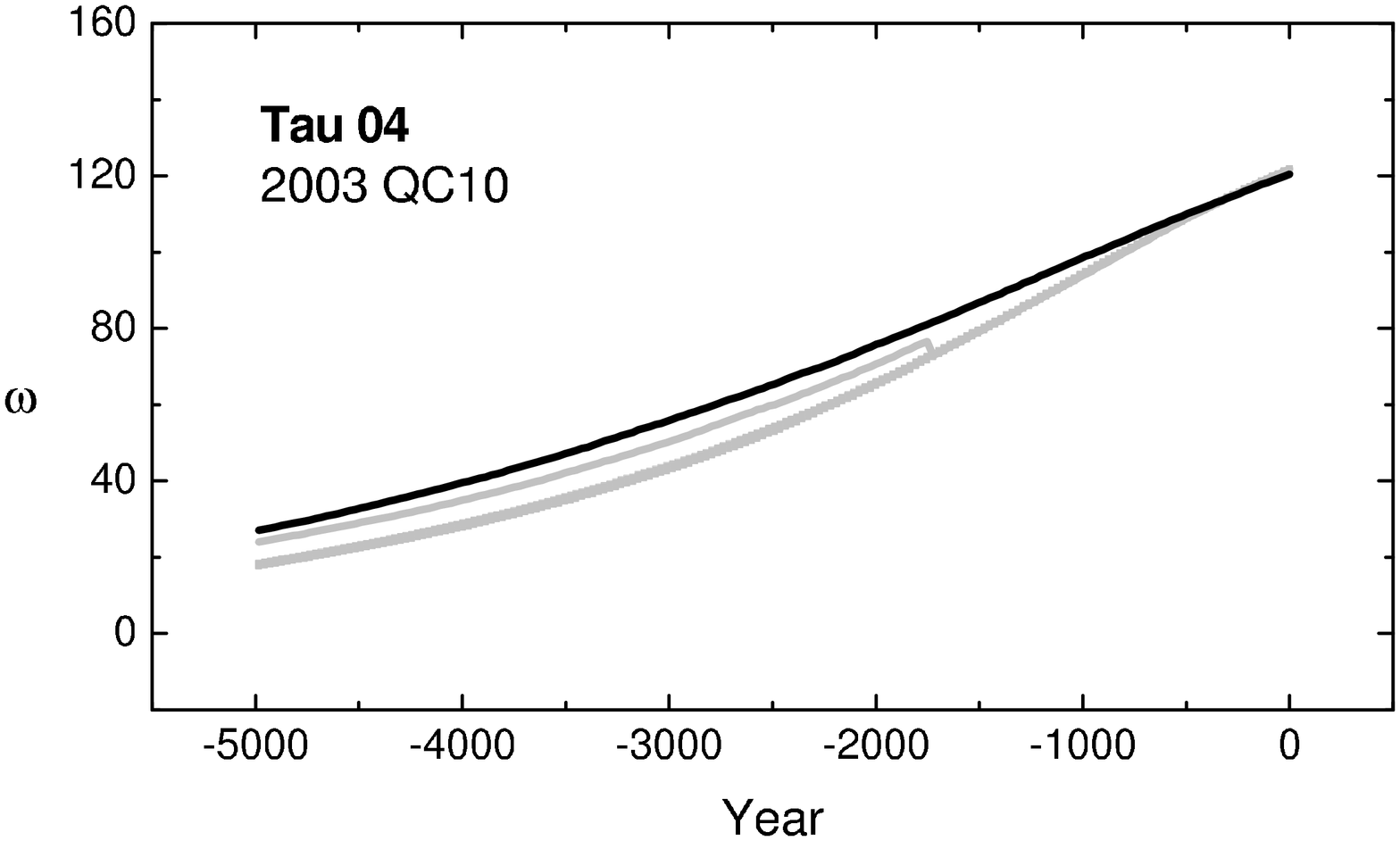}
           }
\vspace{0.7cm}
\centerline{\includegraphics[width=3.3cm,angle=-90]{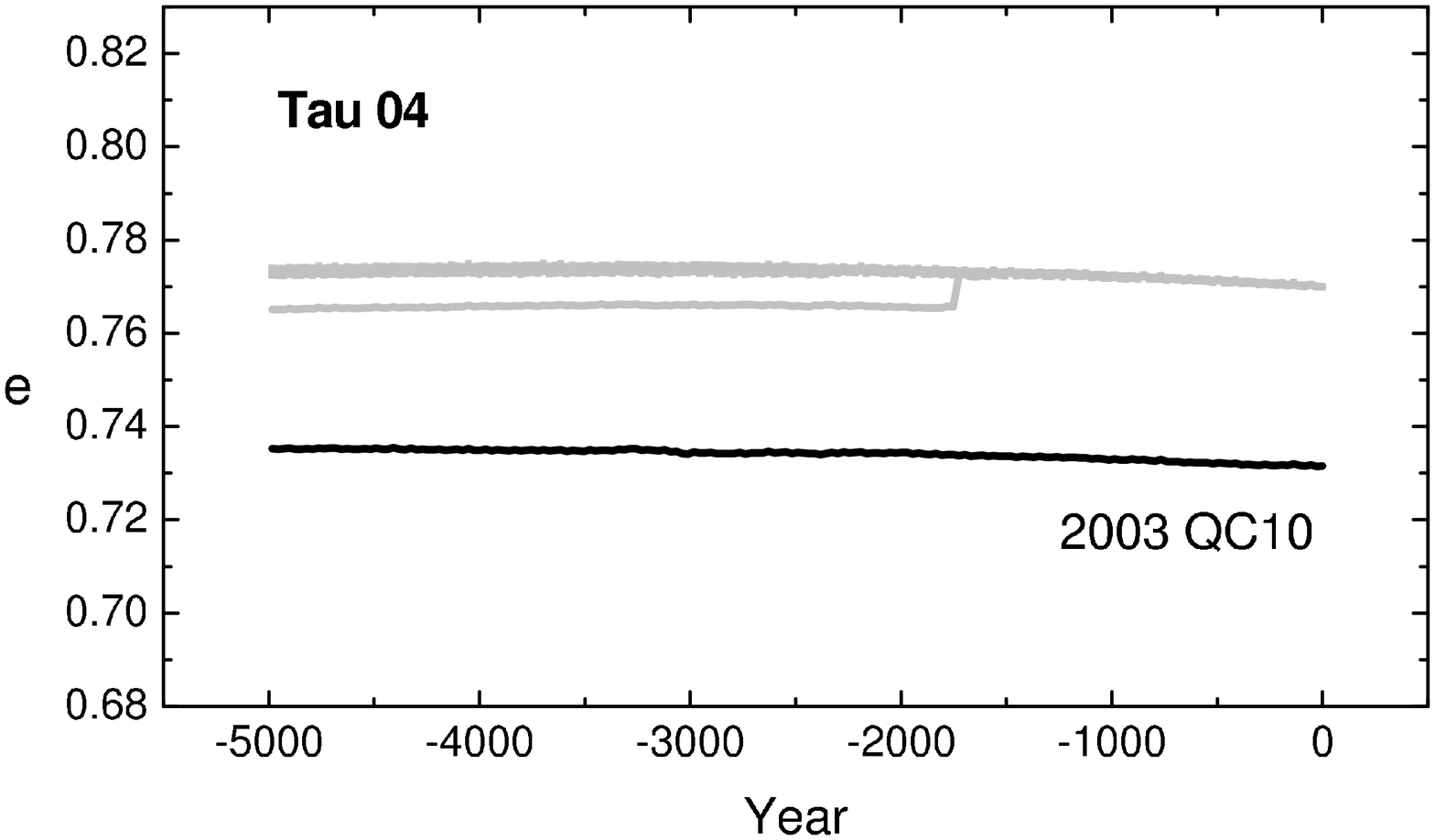}
            \hspace{0.3cm}
            \includegraphics[width=3.3cm,angle=-90]{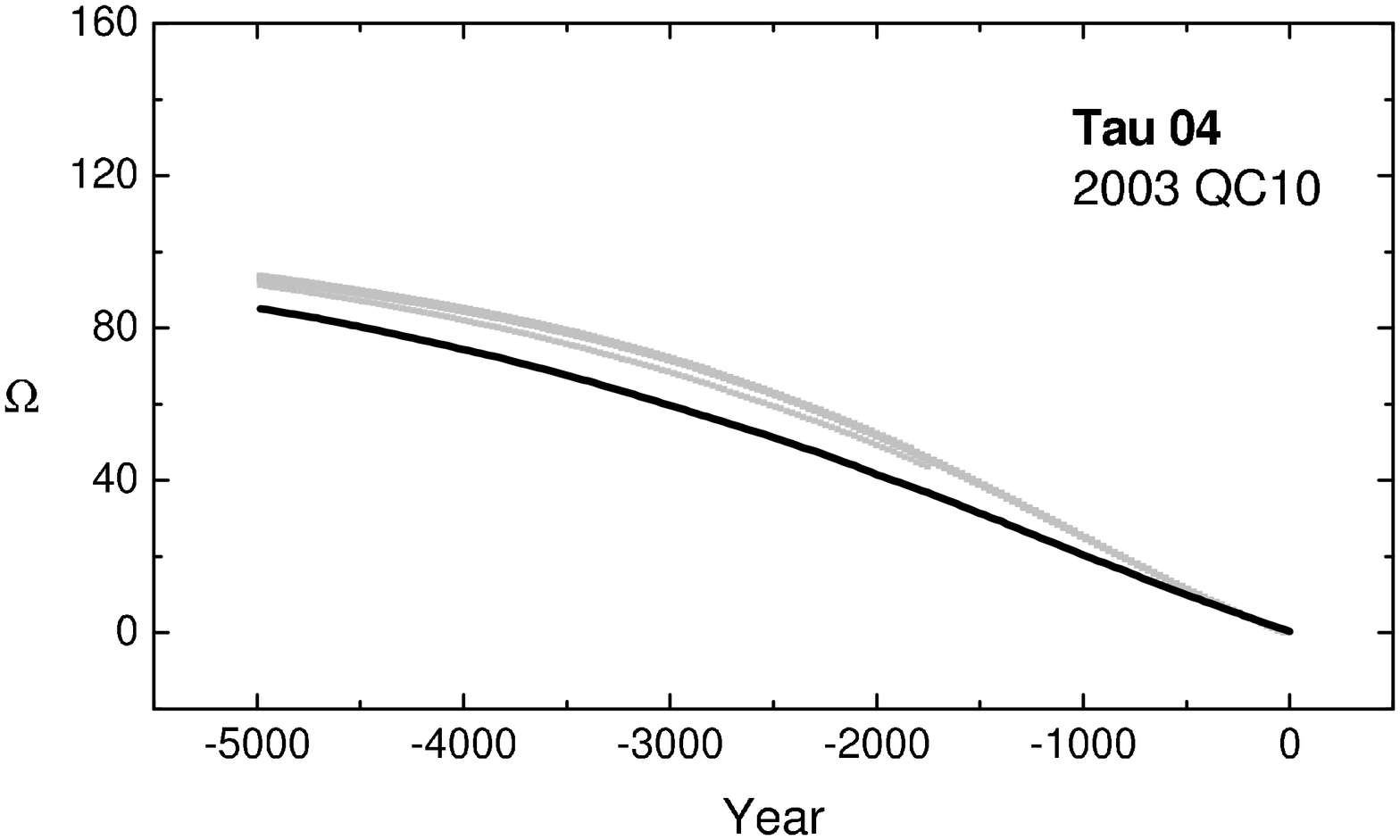}
           }
\vspace{0.7cm}
\centerline{\includegraphics[width=3.3cm,angle=-90]{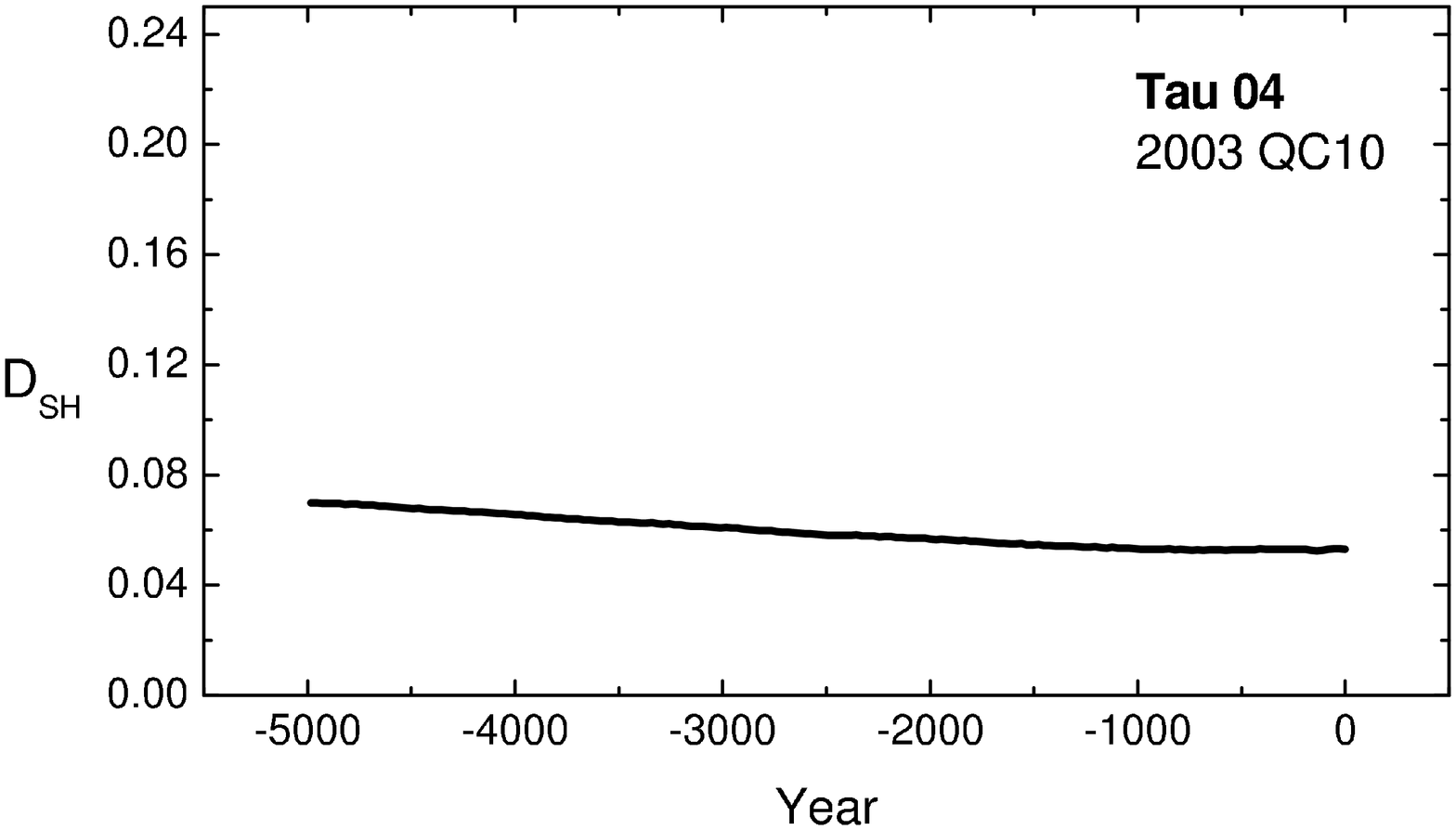}
            \hspace{0.3cm}
            \includegraphics[width=3.3cm,angle=-90]{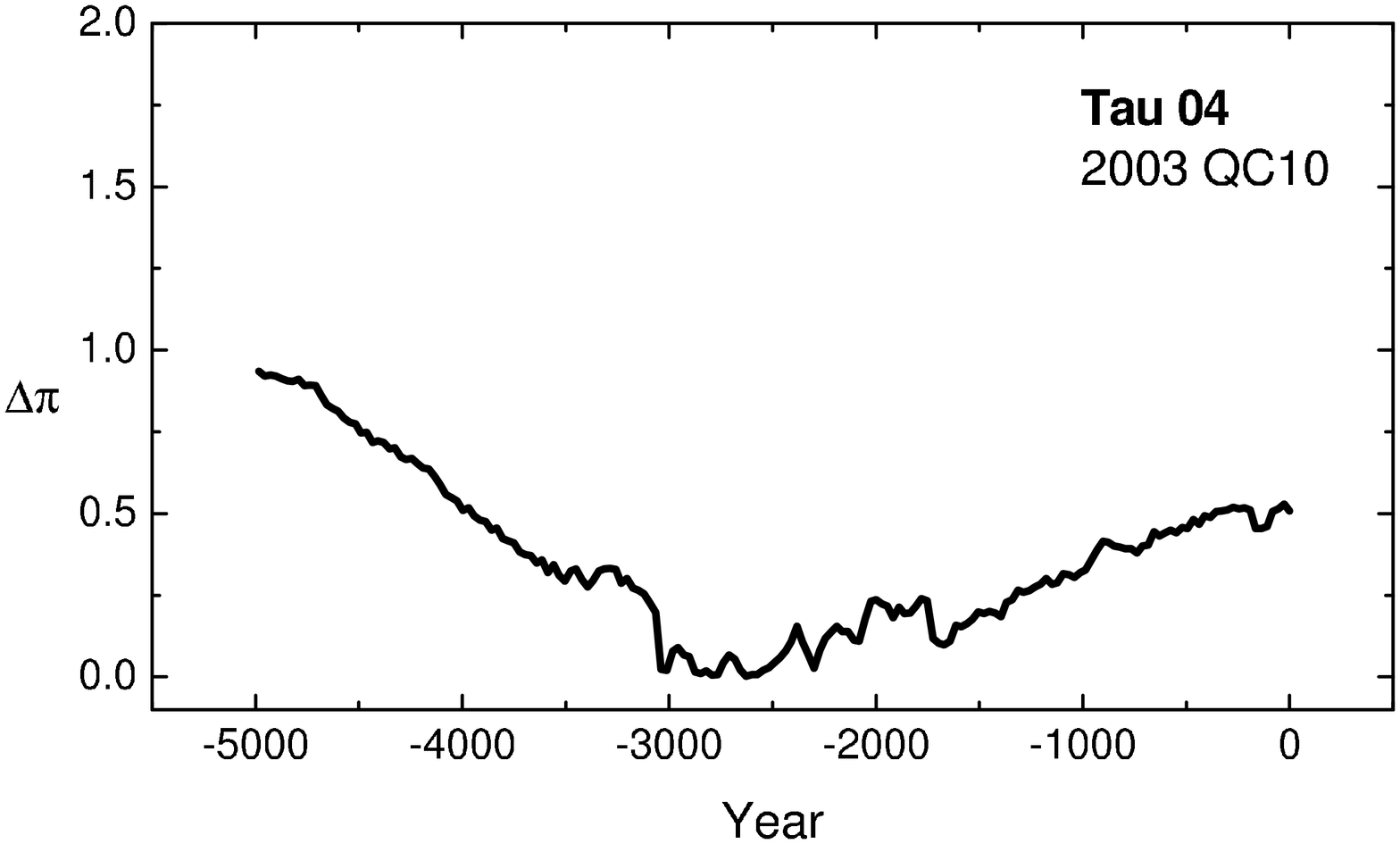}
           }
\caption{Orbital evolution and $D_{SH}$ of TC filament 04 and 2003 QC10.}
\end{figure}
%\clearpage

%Fig. Tau 10
\begin {figure}
\centerline{\includegraphics[width=3.3cm,angle=-90]{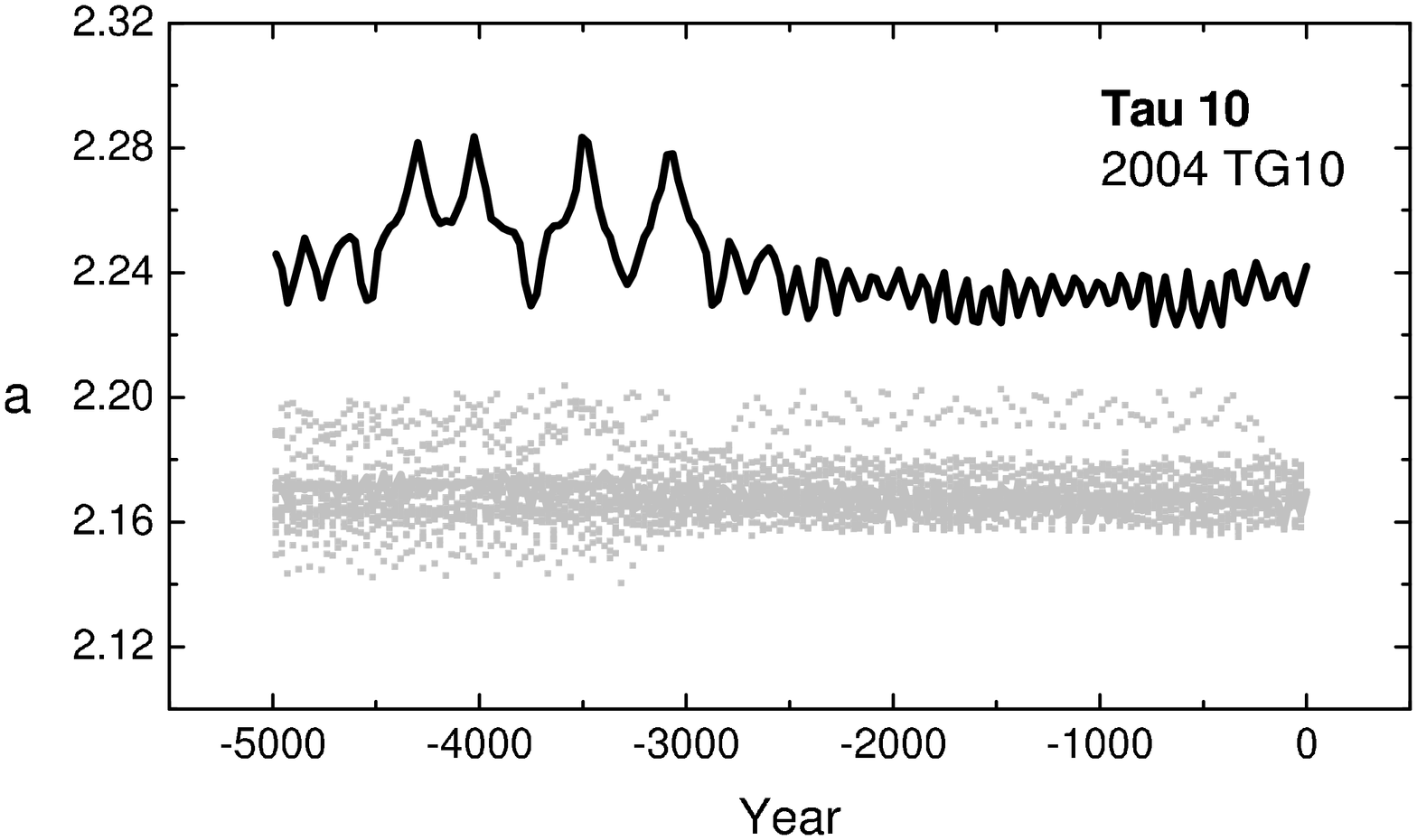}
           \hspace{0.3cm}
            \includegraphics[width=3.3cm,angle=-90]{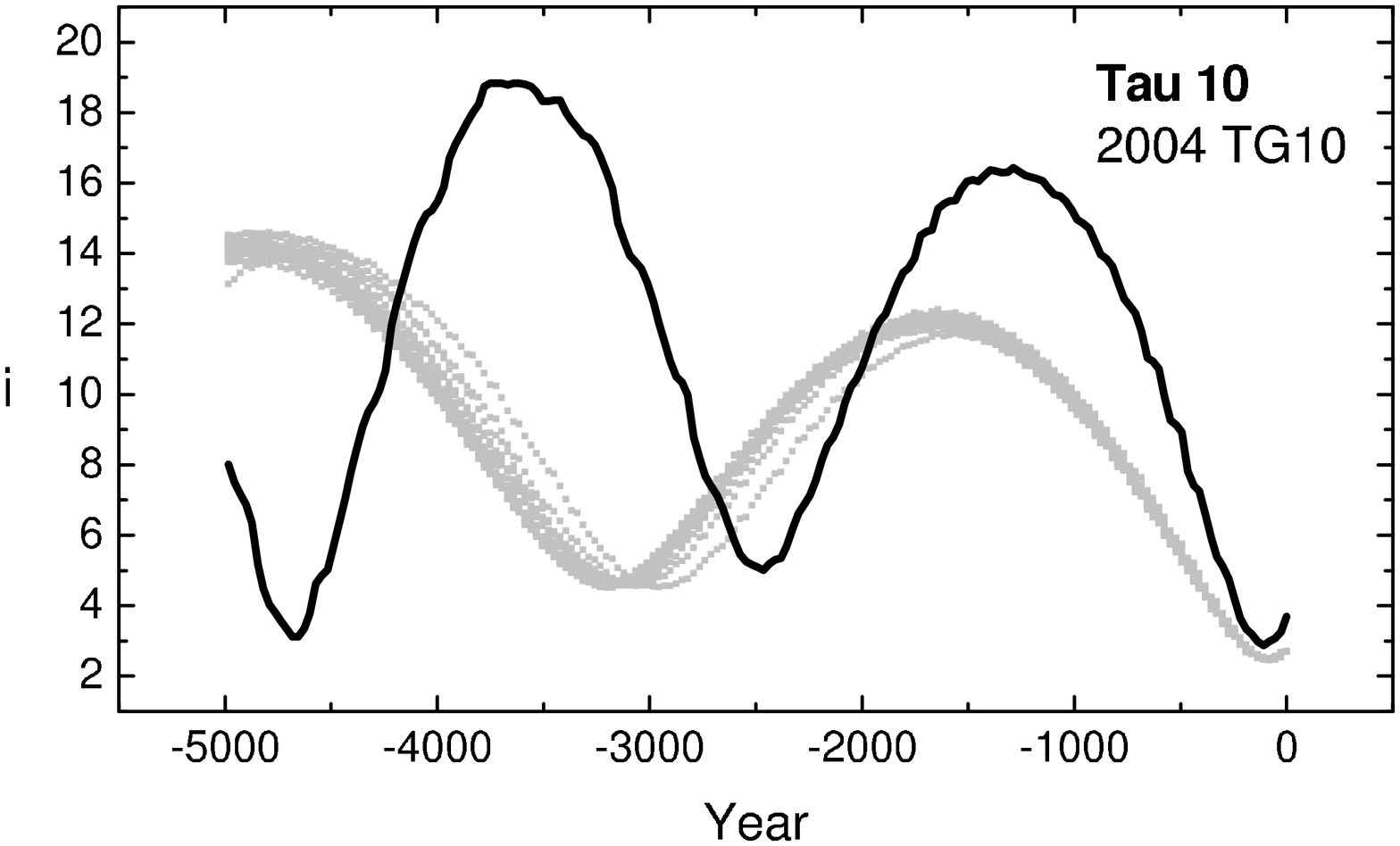}
           }
\vspace{0.7cm}
\centerline{\includegraphics[width=3.3cm,angle=-90]{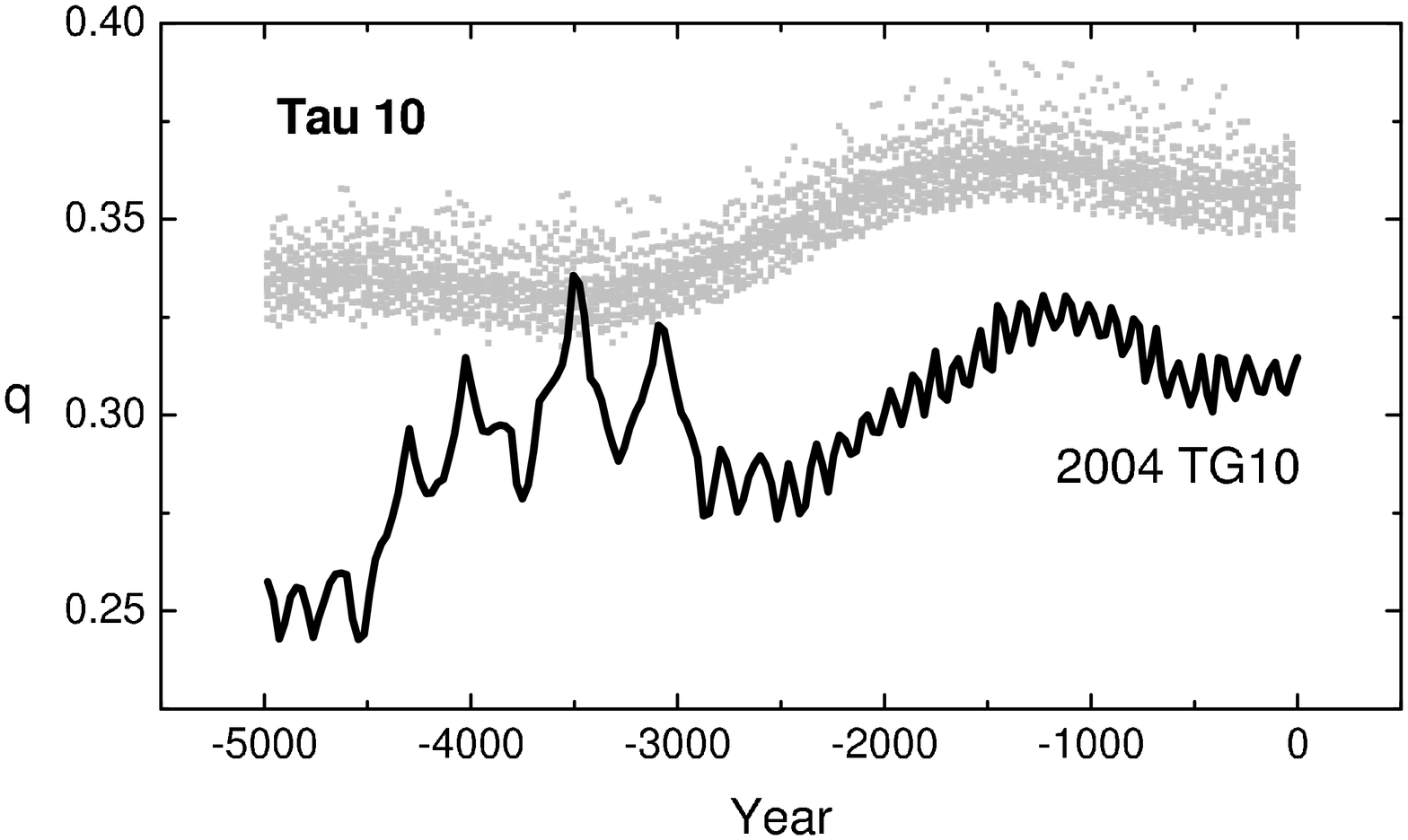}
            \hspace{0.3cm}
            \includegraphics[width=3.3cm,angle=-90]{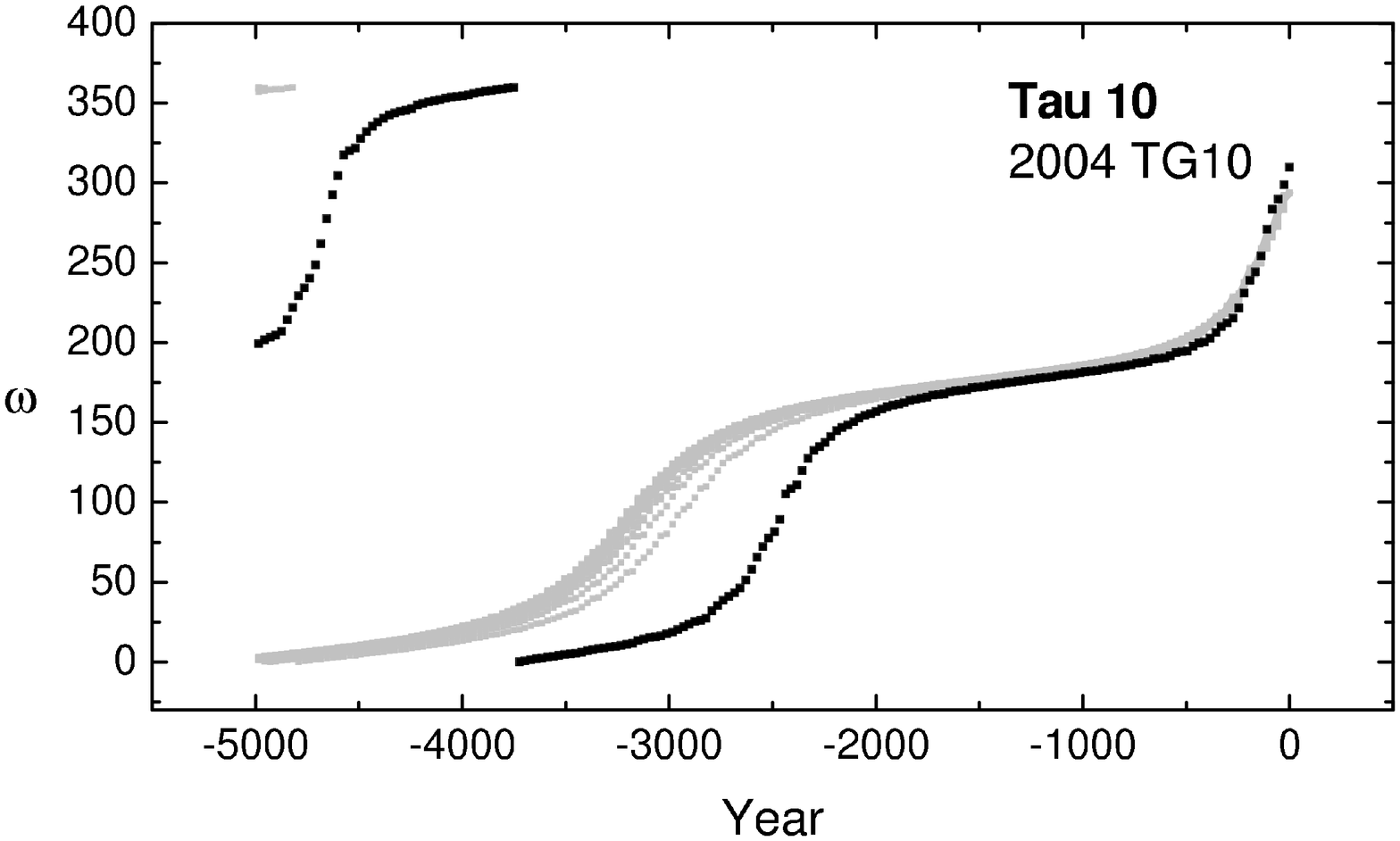}
           }
\vspace{0.7cm}
\centerline{\includegraphics[width=3.3cm,angle=-90]{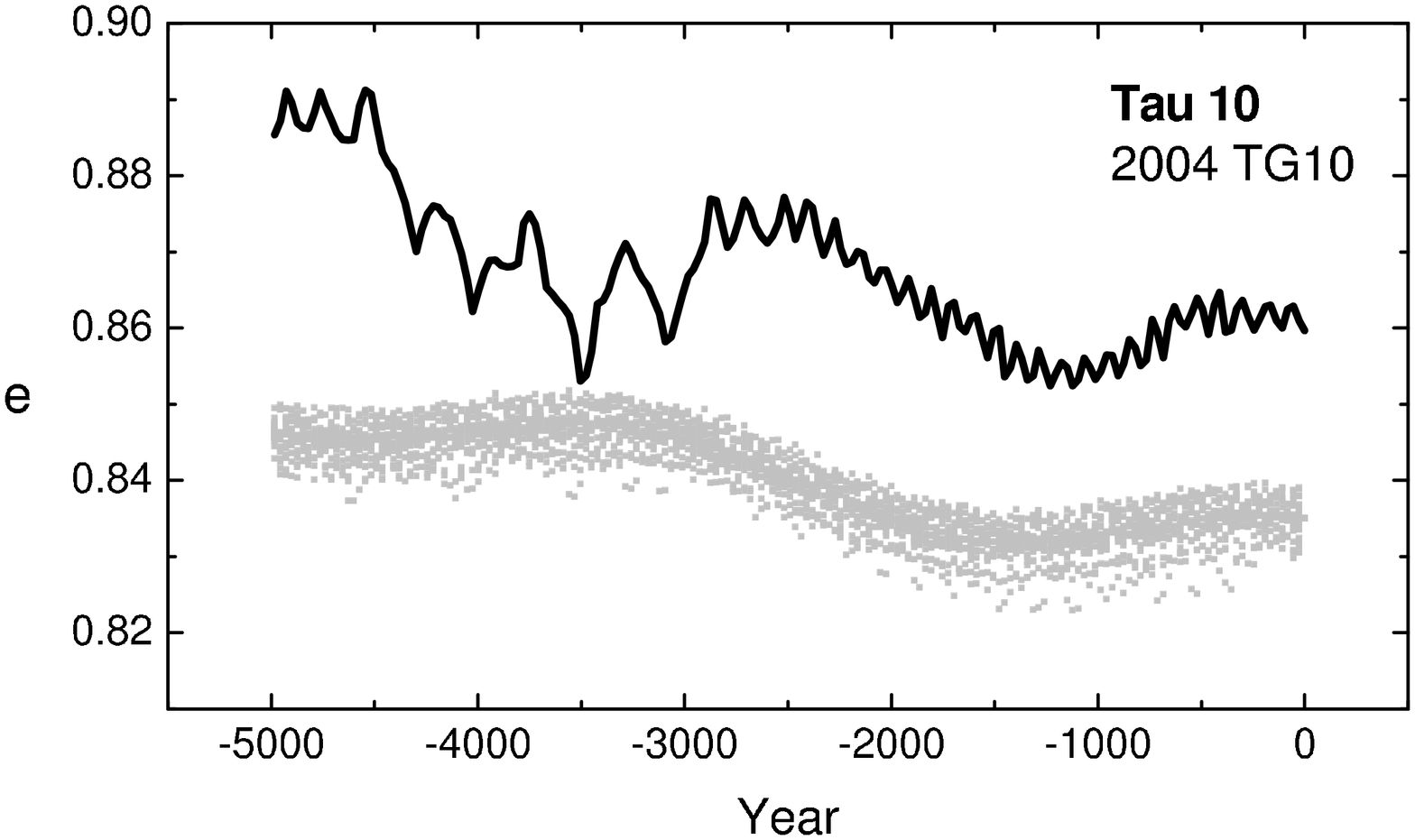}
            \hspace{0.3cm}
            \includegraphics[width=3.3cm,angle=-90]{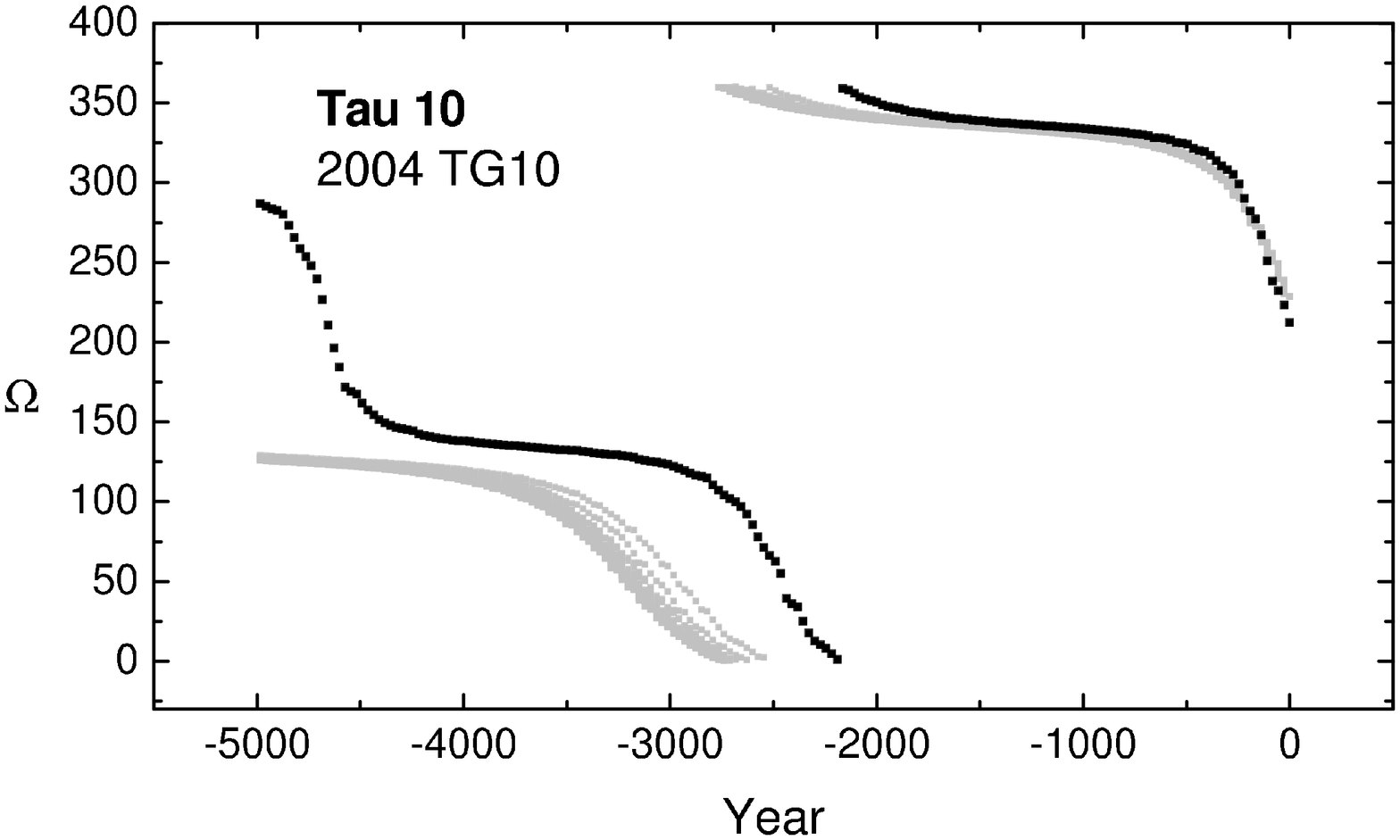}
           }
\vspace{0.7cm}
\centerline{\includegraphics[width=3.3cm,angle=-90]{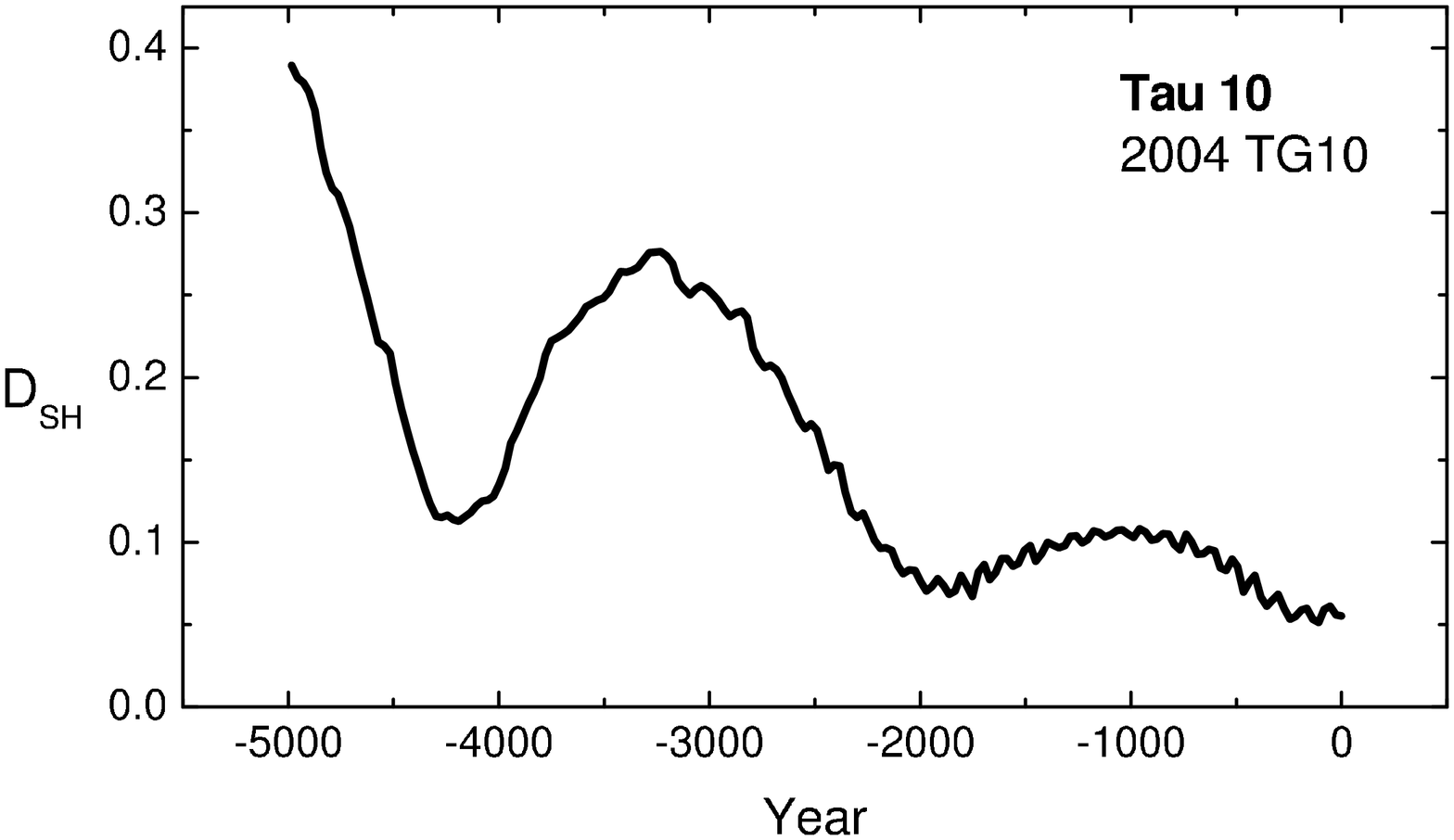}
            \hspace{0.3cm}
            \includegraphics[width=3.3cm,angle=-90]{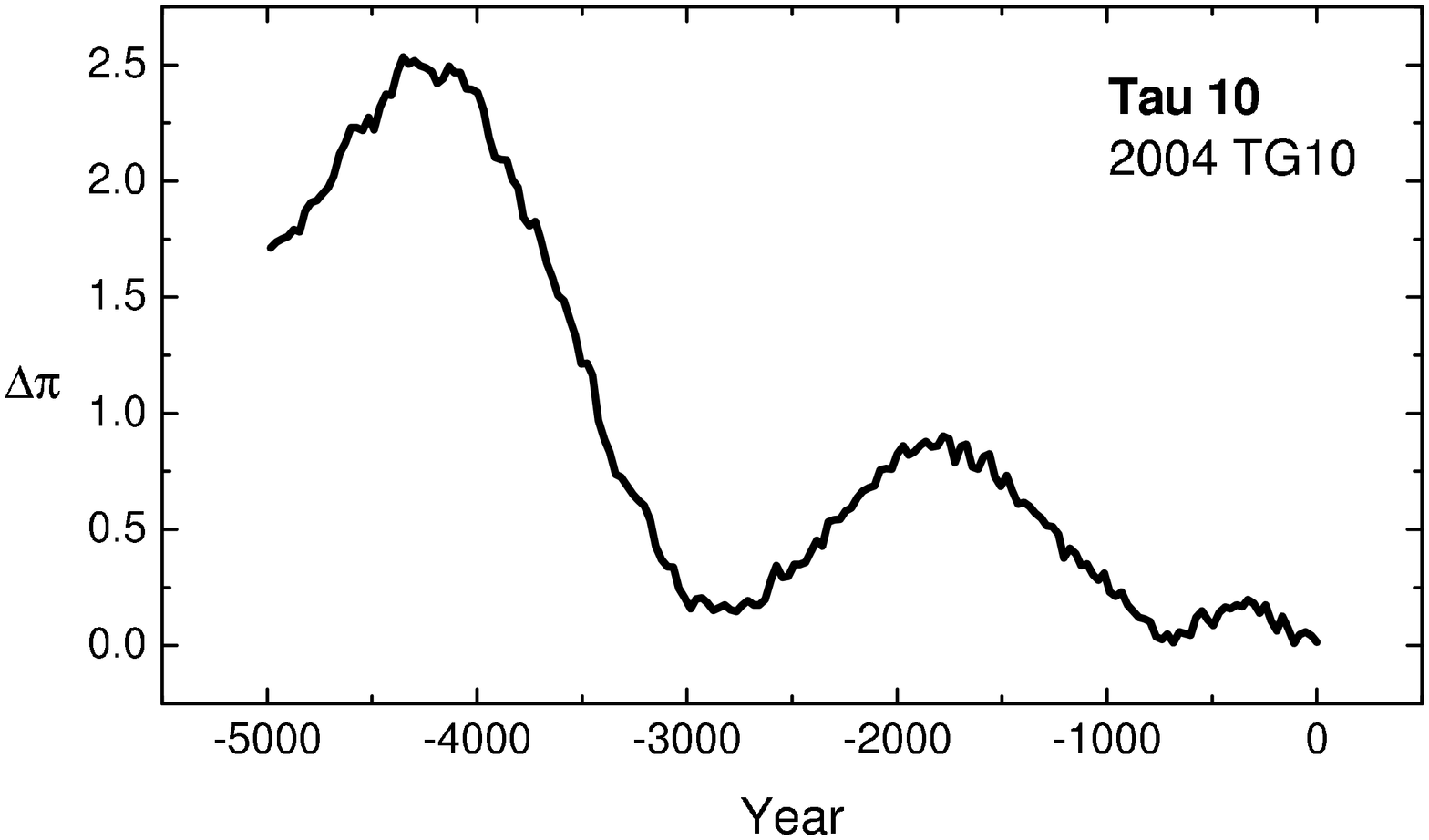}
           }
\caption{Orbital evolution and $D_{SH}$ of TC filament 10 and 2004 TG10.}
\end{figure}
%\clearpage

%Fig. Tau 11
\begin {figure}
\centerline{\includegraphics[width=3.3cm,angle=-90]{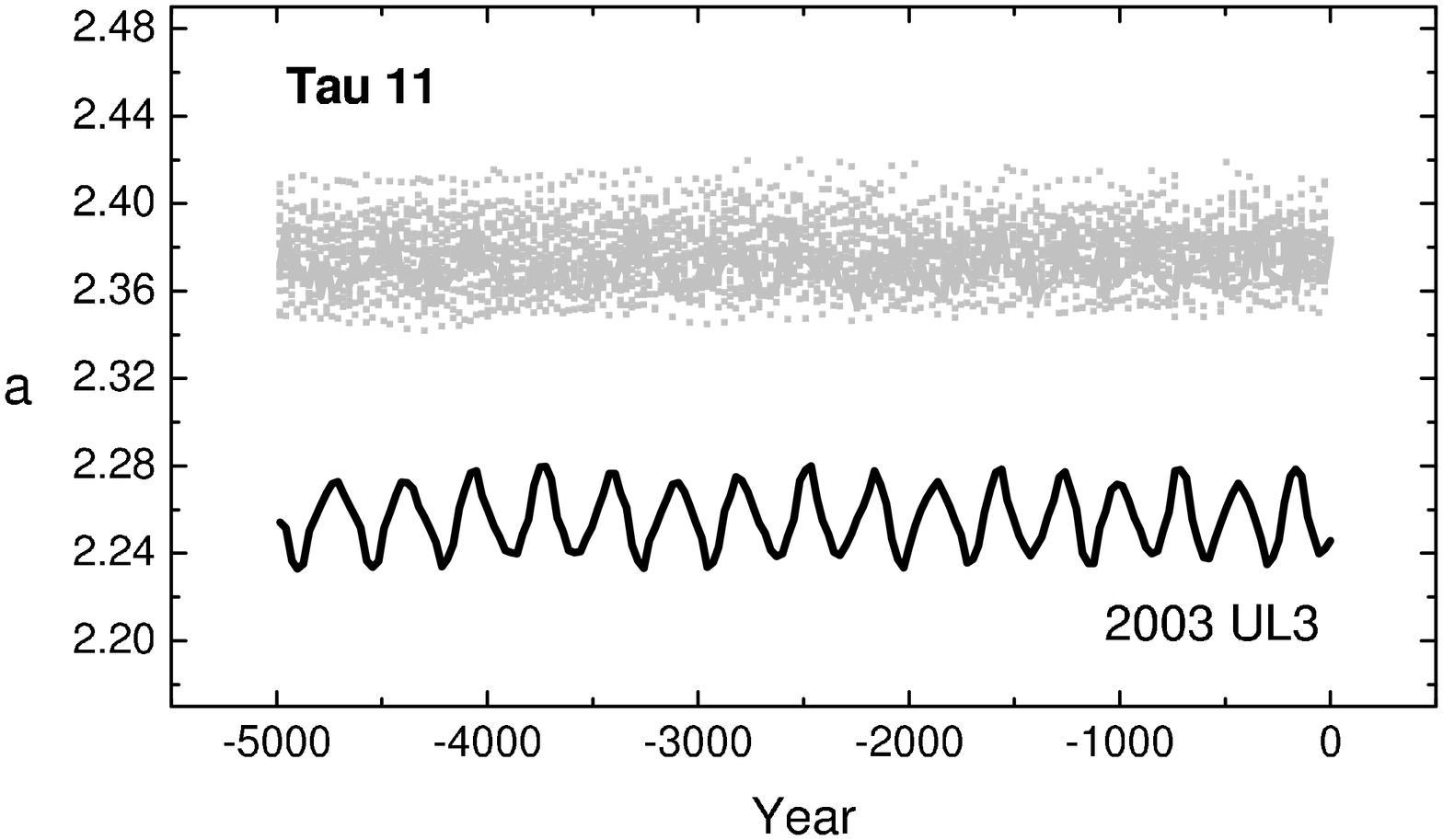}
           \hspace{0.3cm}
            \includegraphics[width=3.3cm,angle=-90]{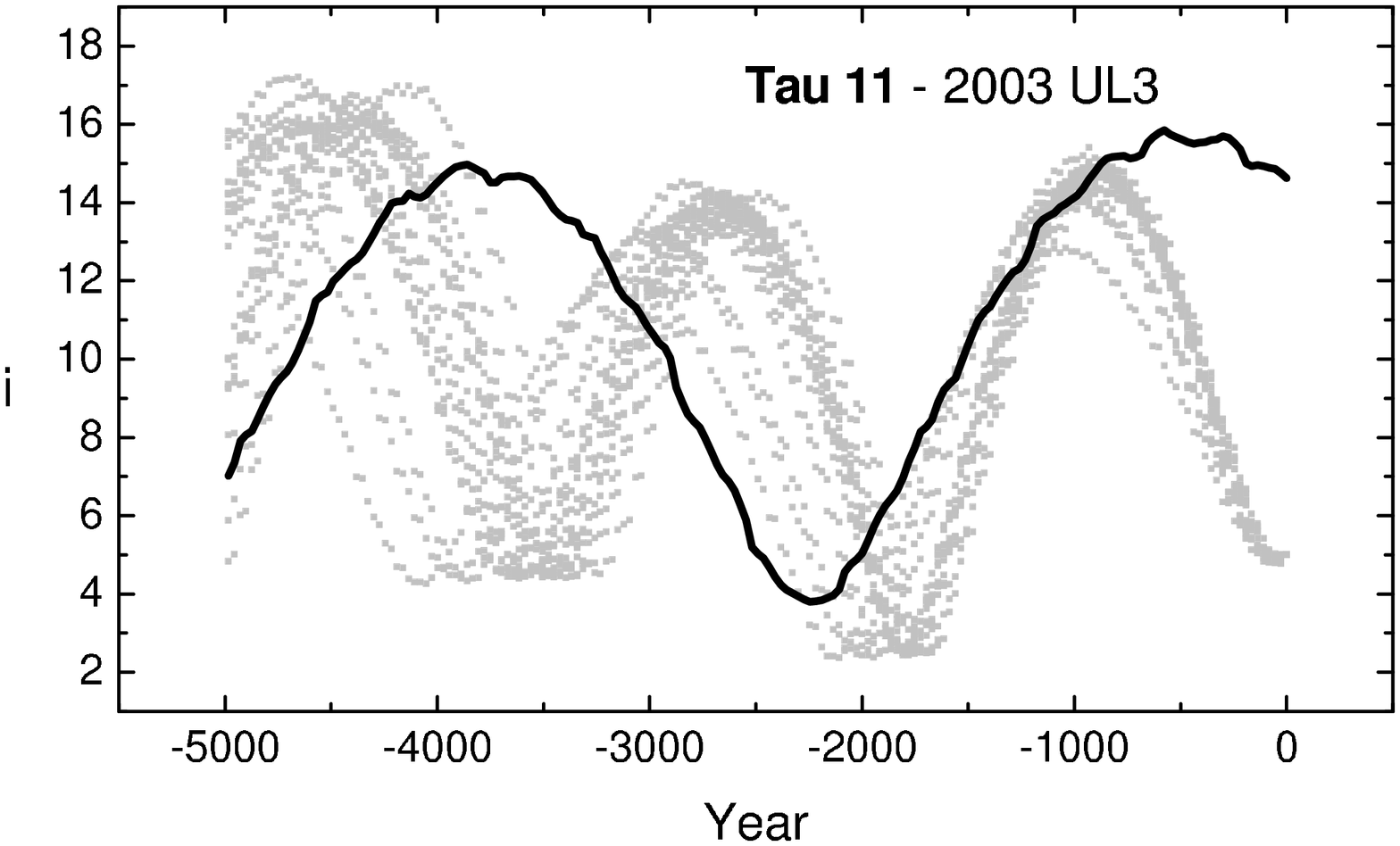}
           }
\vspace{0.7cm}
\centerline{\includegraphics[width=3.3cm,angle=-90]{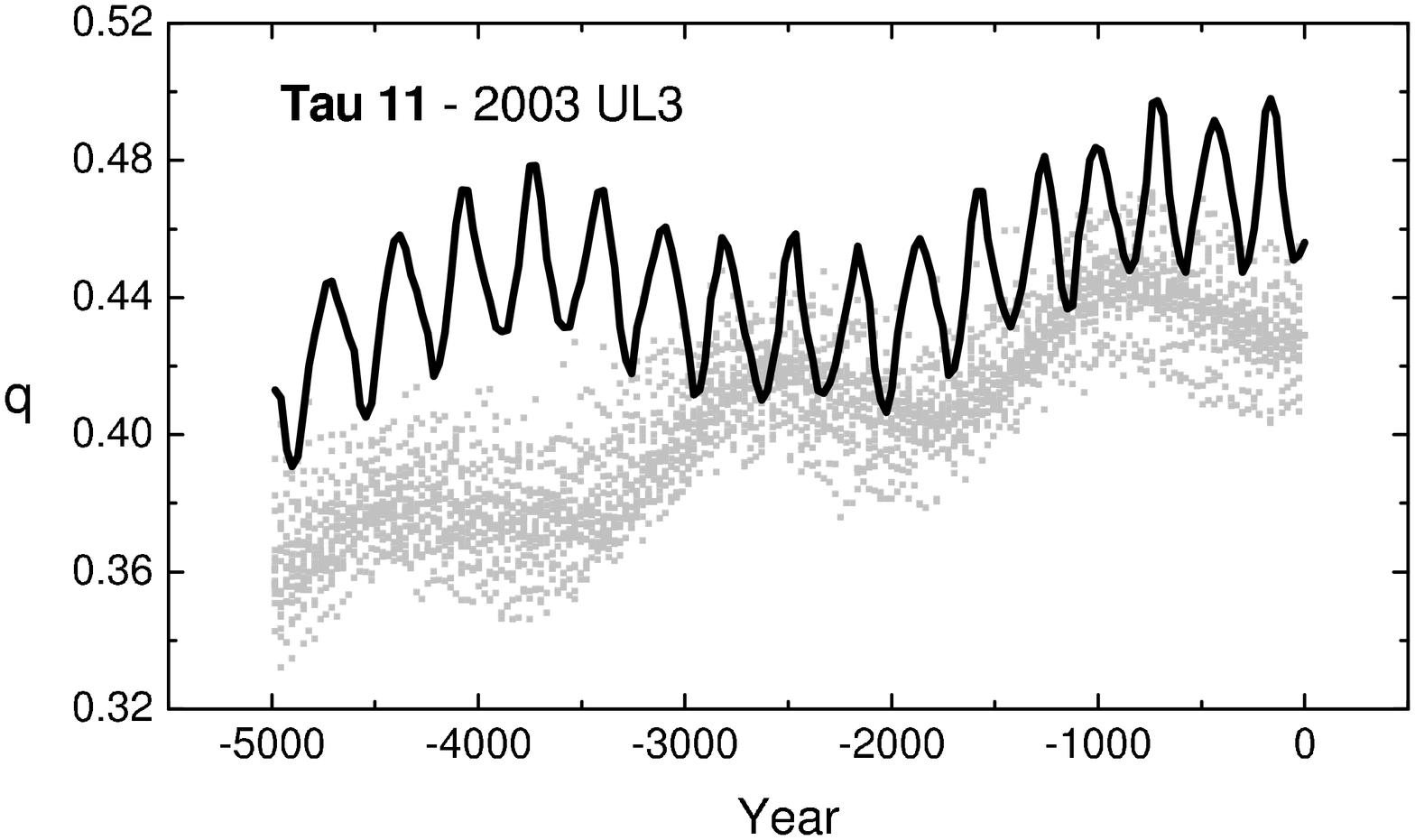}
            \hspace{0.3cm}
            \includegraphics[width=3.3cm,angle=-90]{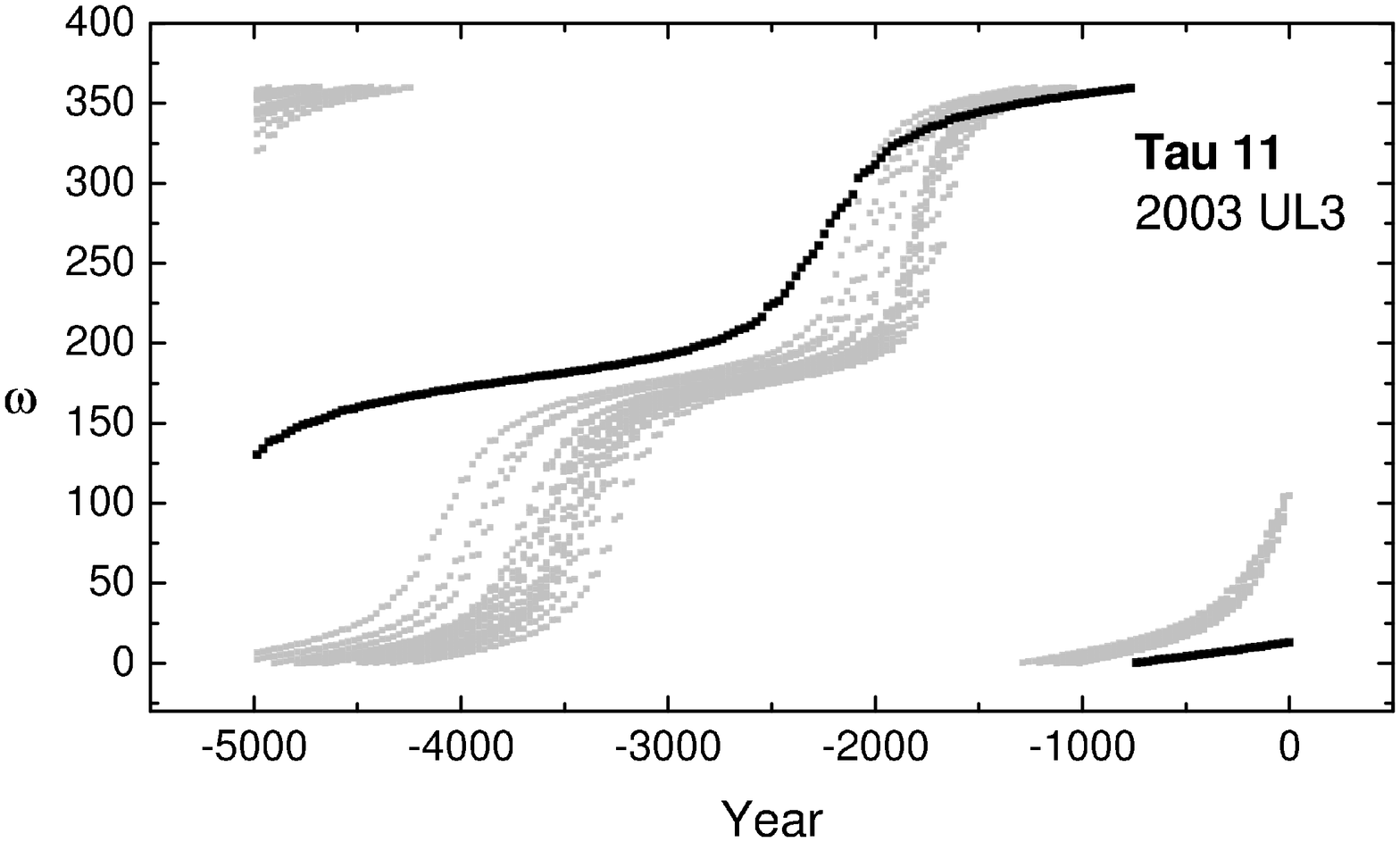}
           }
\vspace{0.7cm}
\centerline{\includegraphics[width=3.3cm,angle=-90]{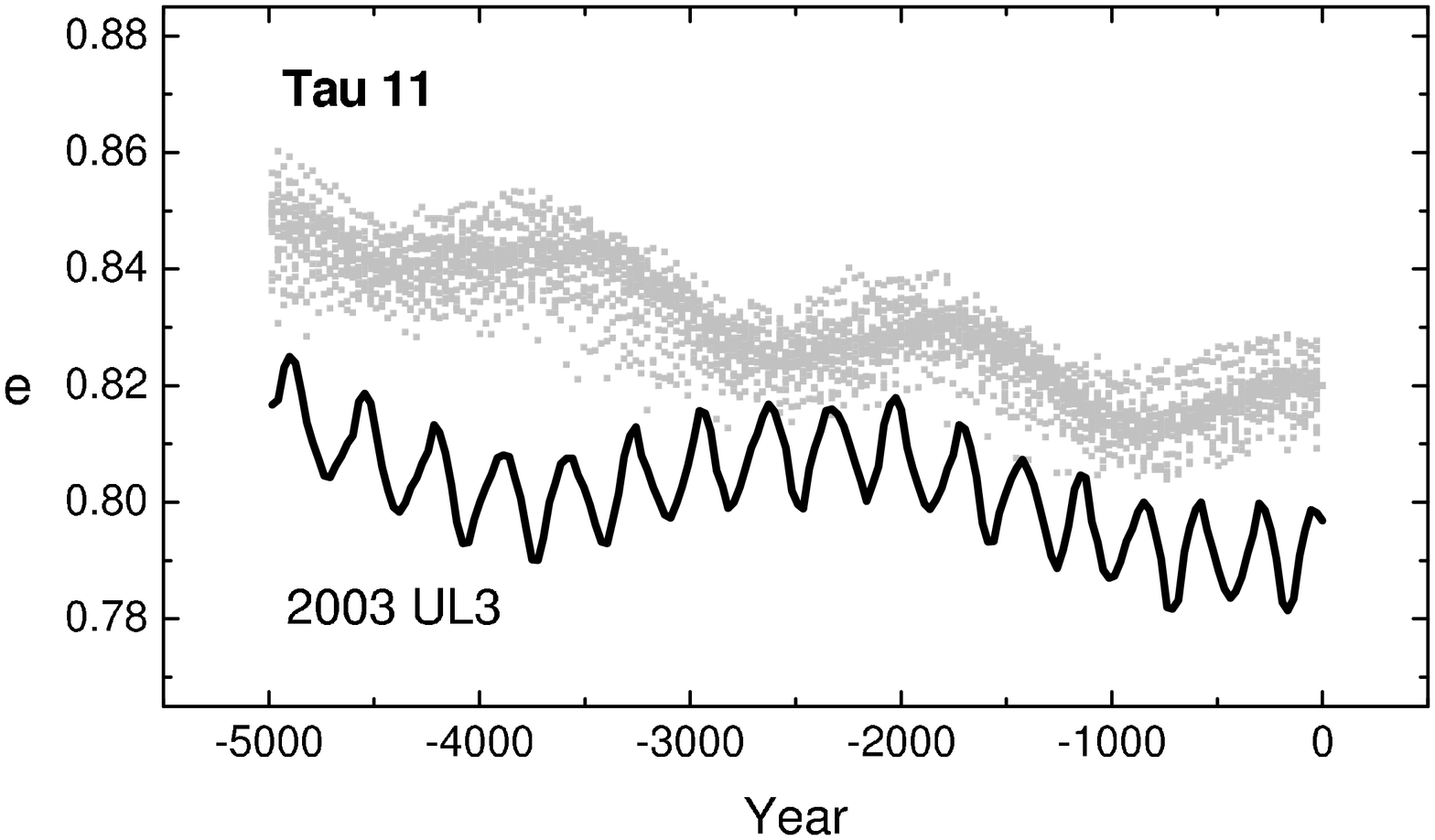}
            \hspace{0.3cm}
            \includegraphics[width=3.3cm,angle=-90]{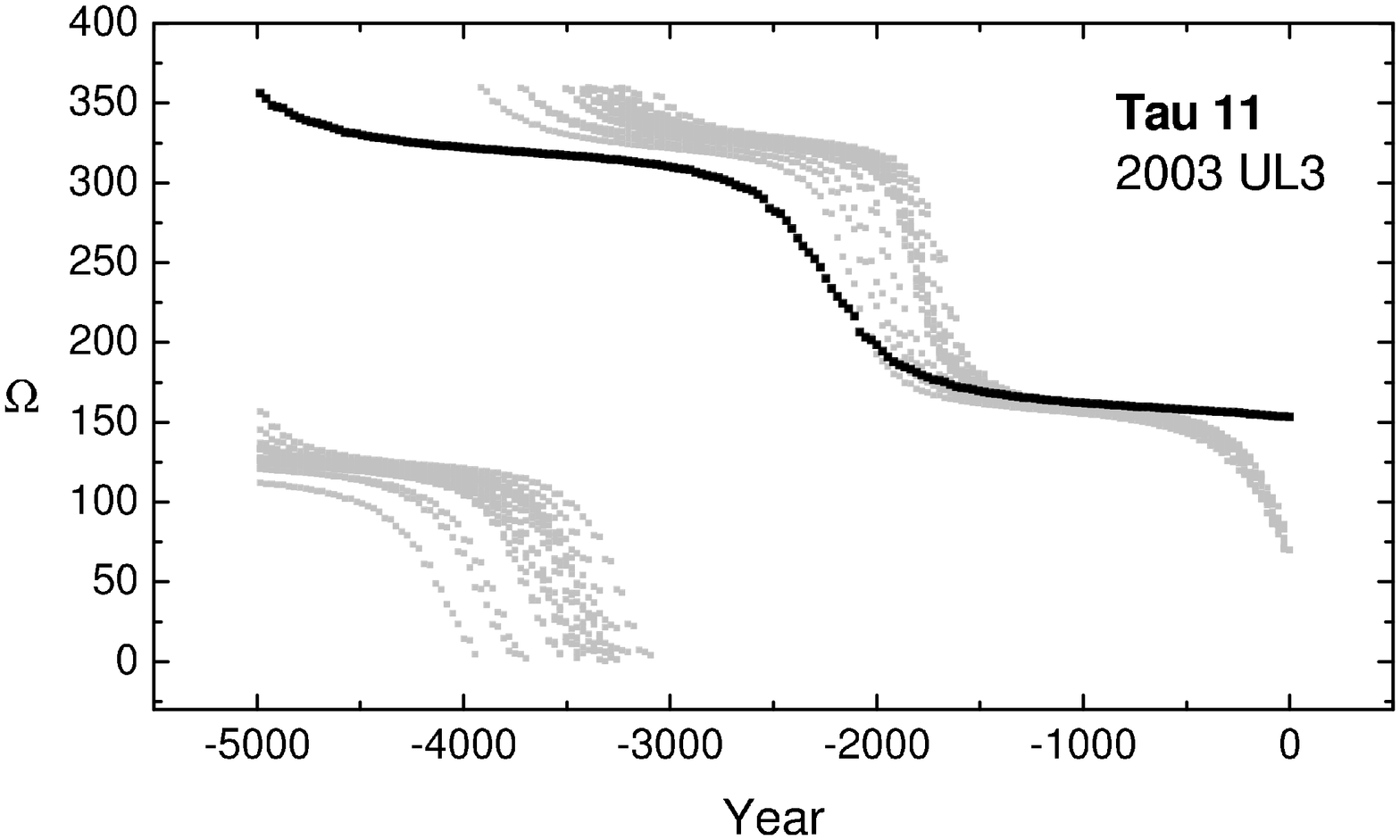}
           }
\vspace{0.7cm}
\centerline{\includegraphics[width=3.3cm,angle=-90]{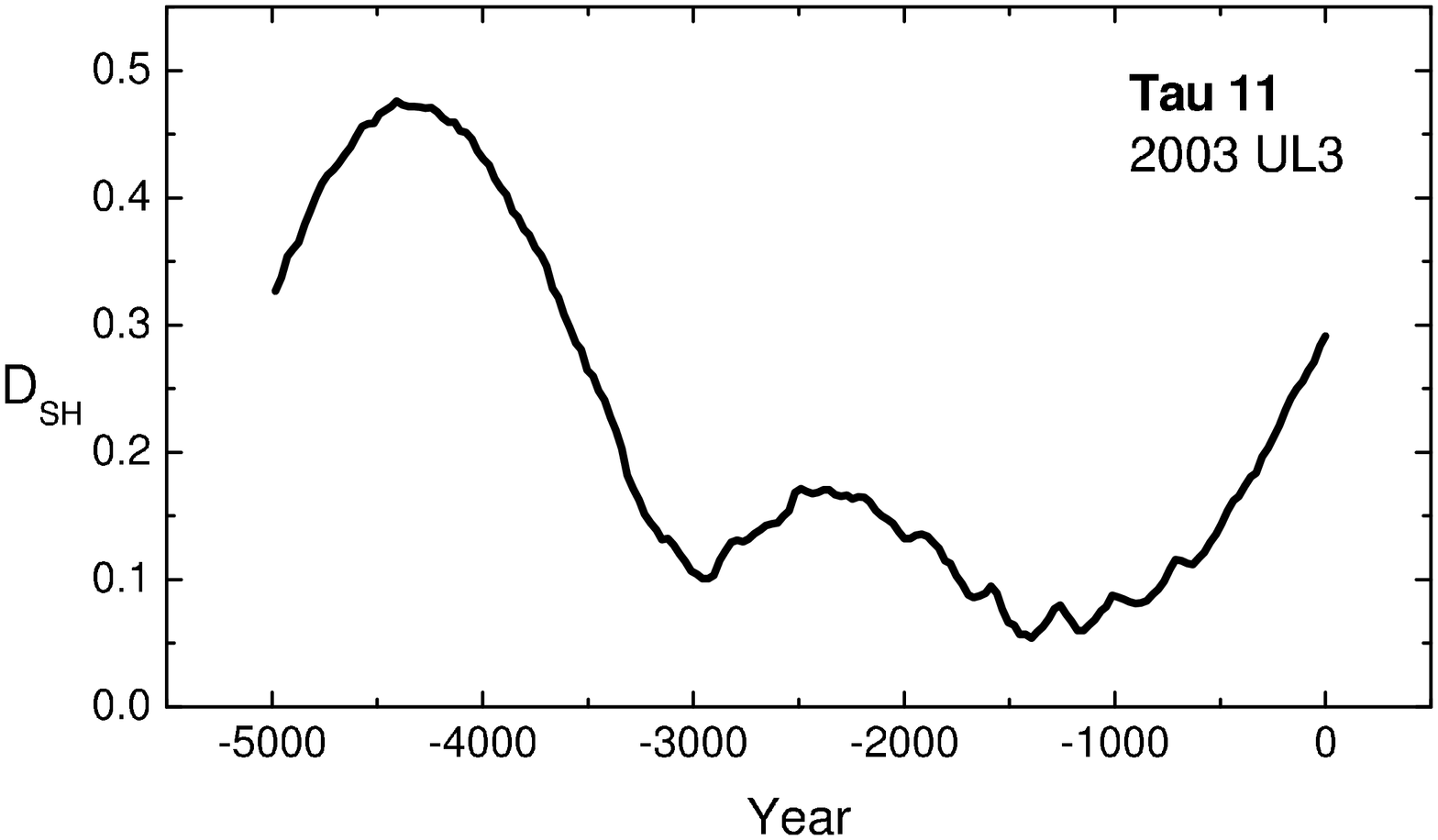}
            \hspace{0.3cm}
            \includegraphics[width=3.3cm,angle=-90]{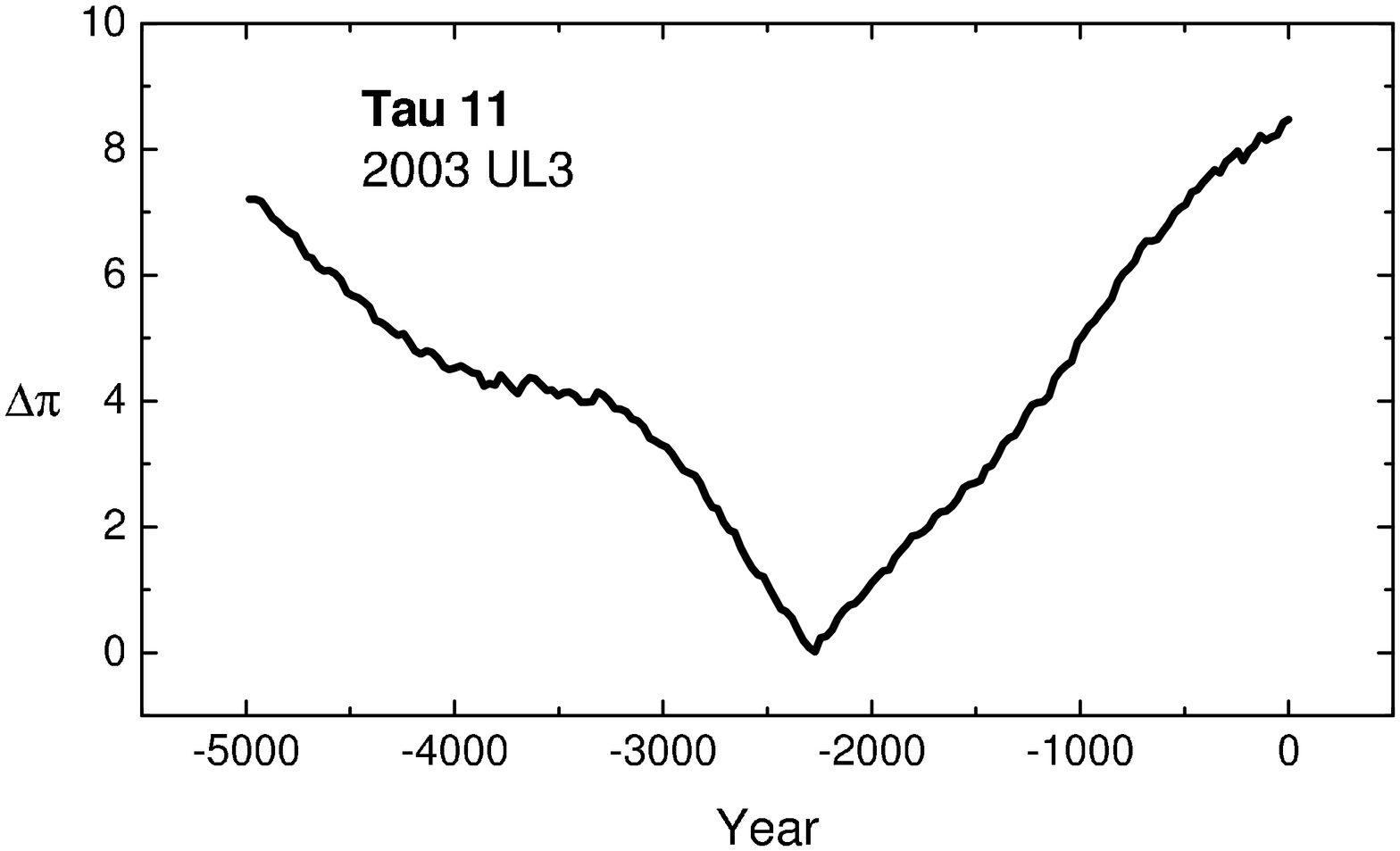}
           }
\caption{Orbital evolution and $D_{SH}$ of TC filament 11 and 2003 UL3.}
\end{figure}

%Fig. Tau 12
\begin {figure}
\centerline{\includegraphics[width=3.3cm,angle=-90]{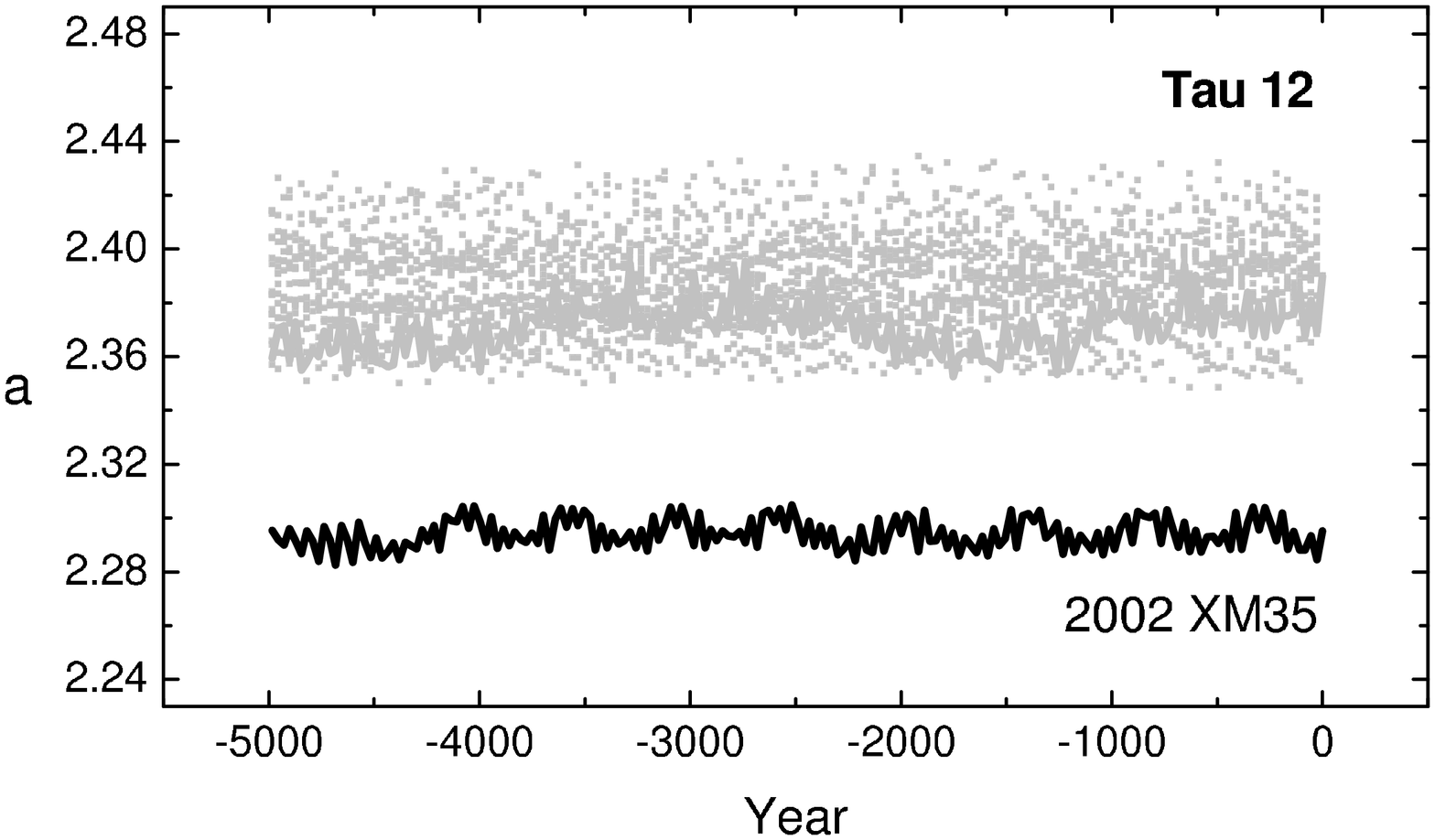}
           \hspace{0.3cm}
            \includegraphics[width=3.3cm,angle=-90]{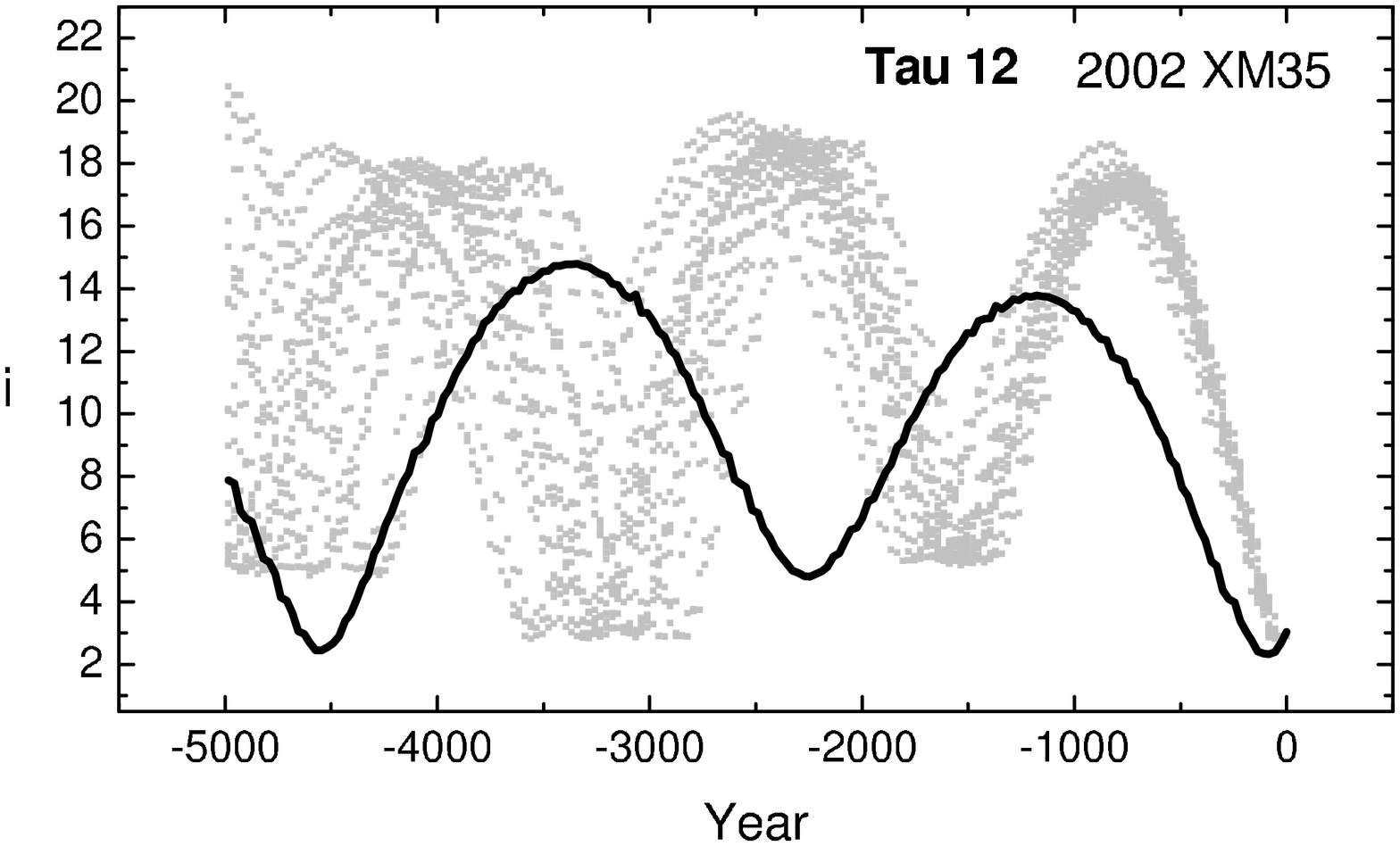}
           }
\vspace{0.7cm}
\centerline{\includegraphics[width=3.3cm,angle=-90]{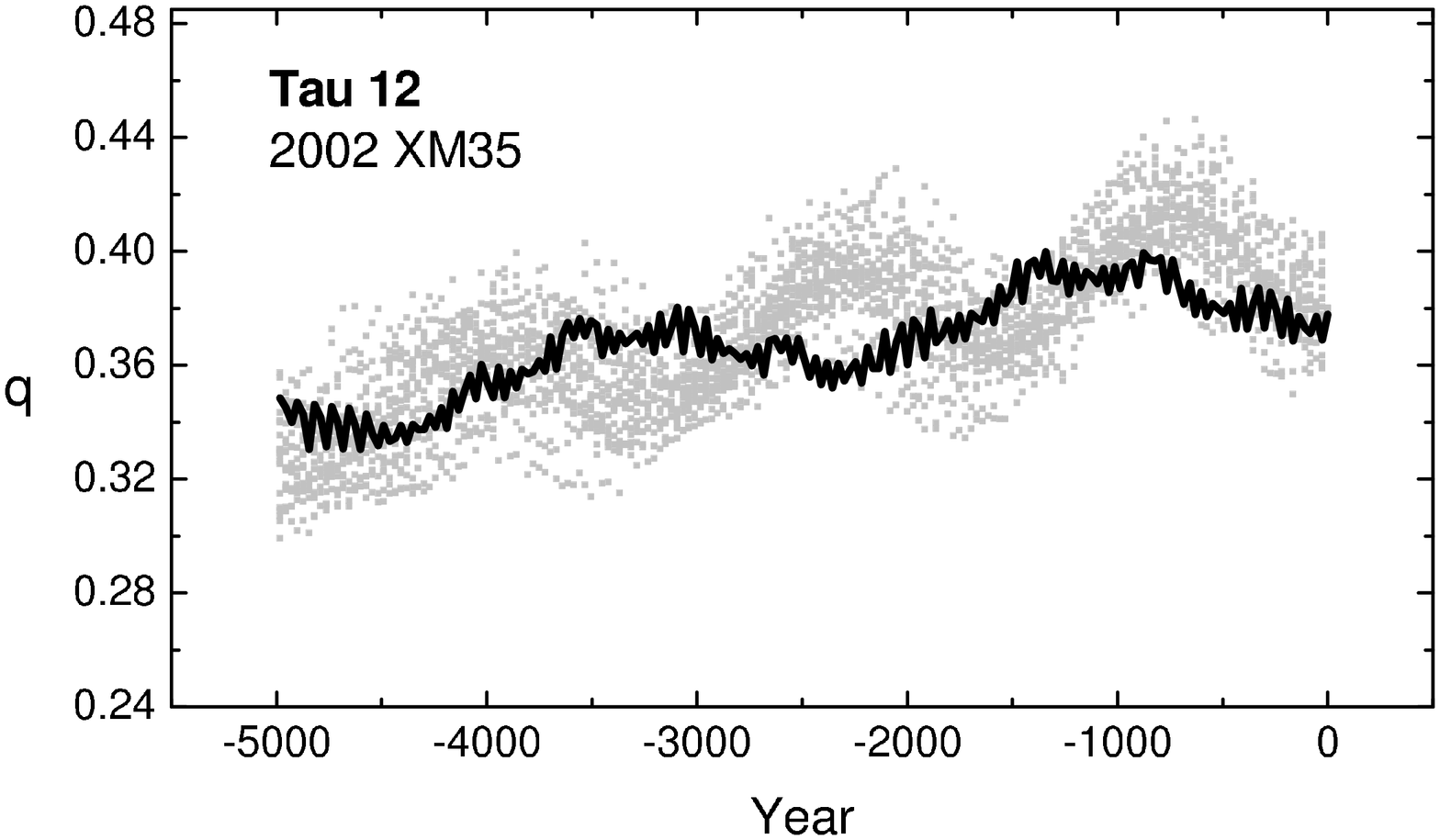}
            \hspace{0.3cm}
            \includegraphics[width=3.3cm,angle=-90]{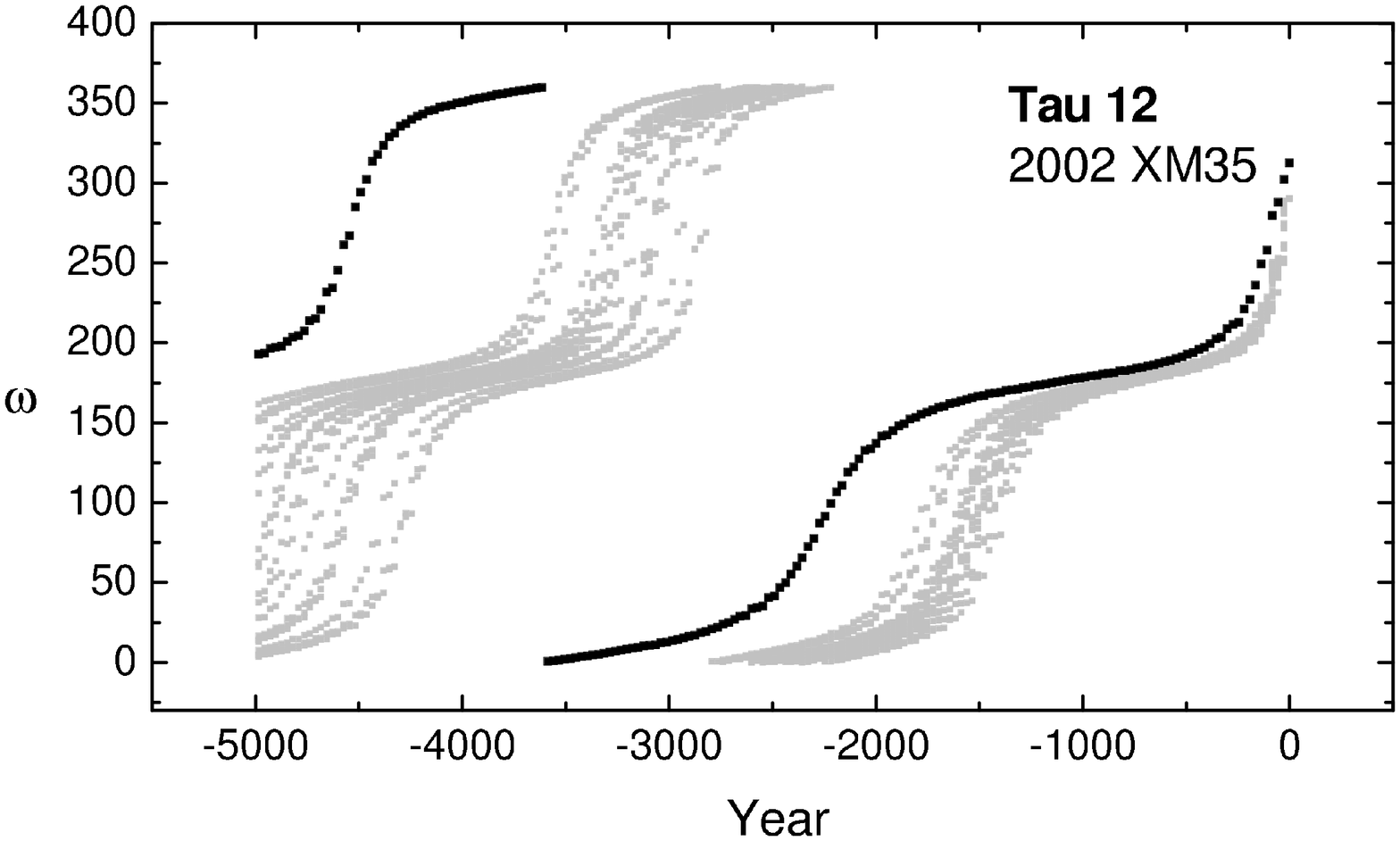}
           }
\vspace{0.7cm}
\centerline{\includegraphics[width=3.3cm,angle=-90]{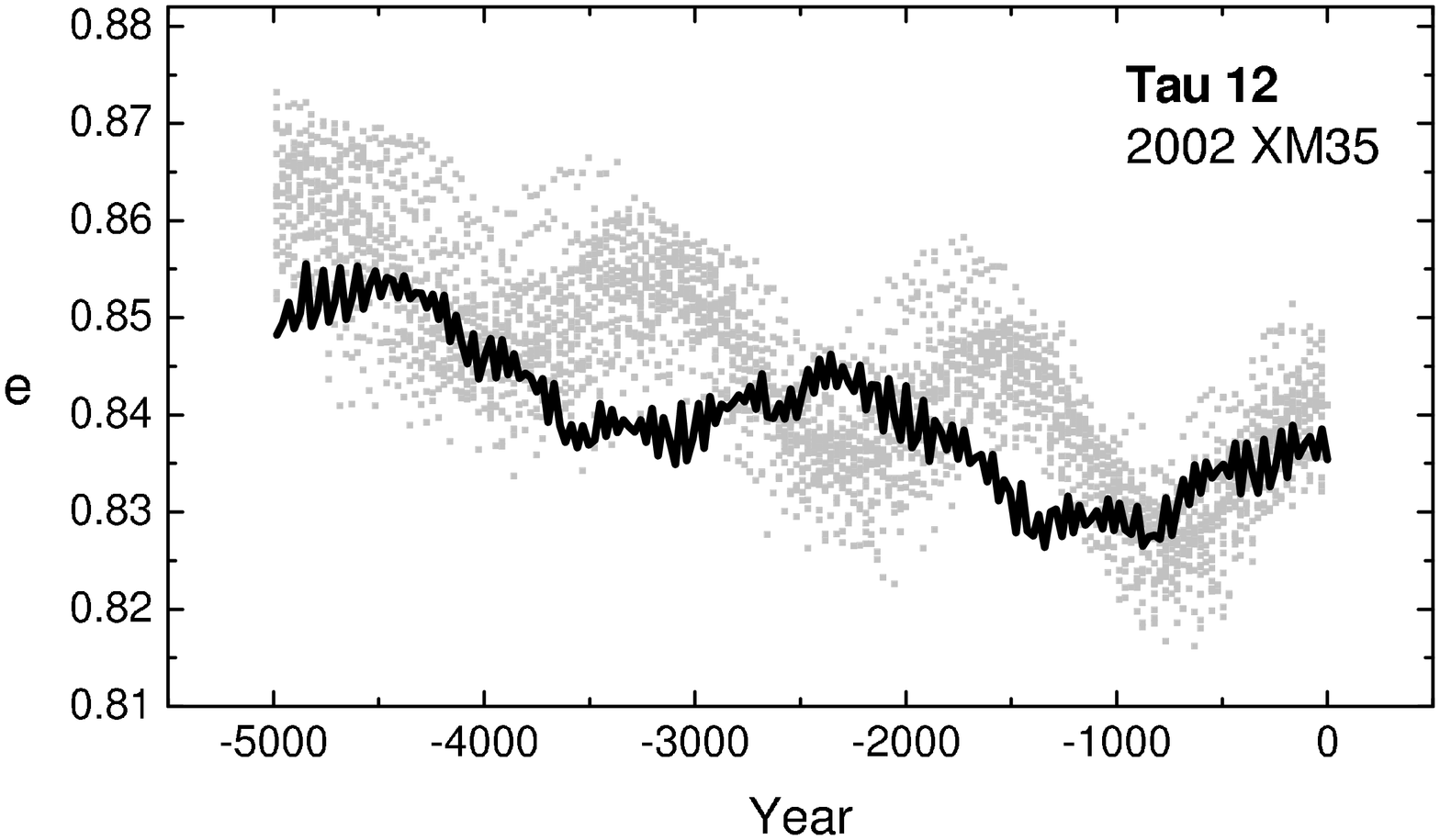}
            \hspace{0.3cm}
            \includegraphics[width=3.3cm,angle=-90]{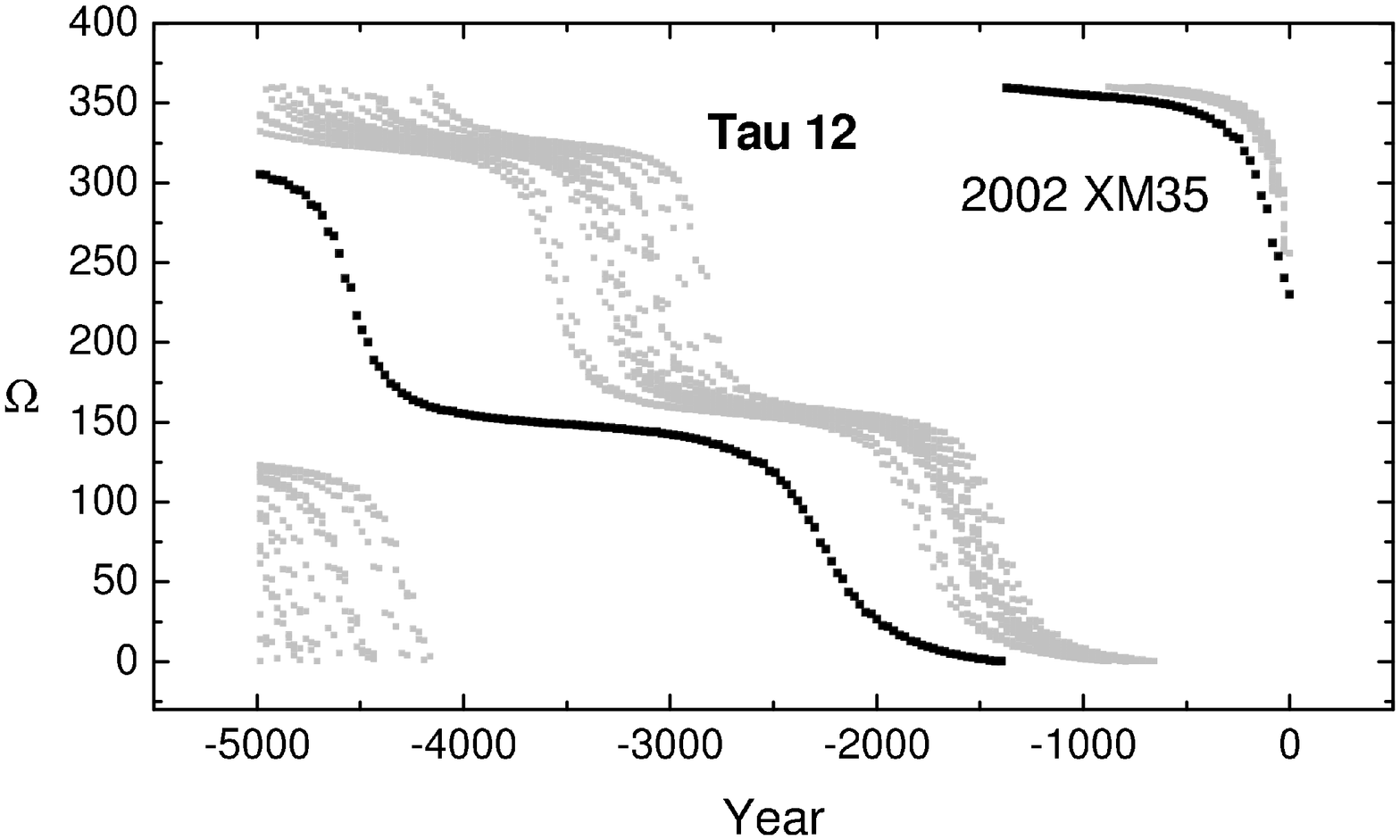}
           }
\vspace{0.7cm}
\centerline{\includegraphics[width=3.3cm,angle=-90]{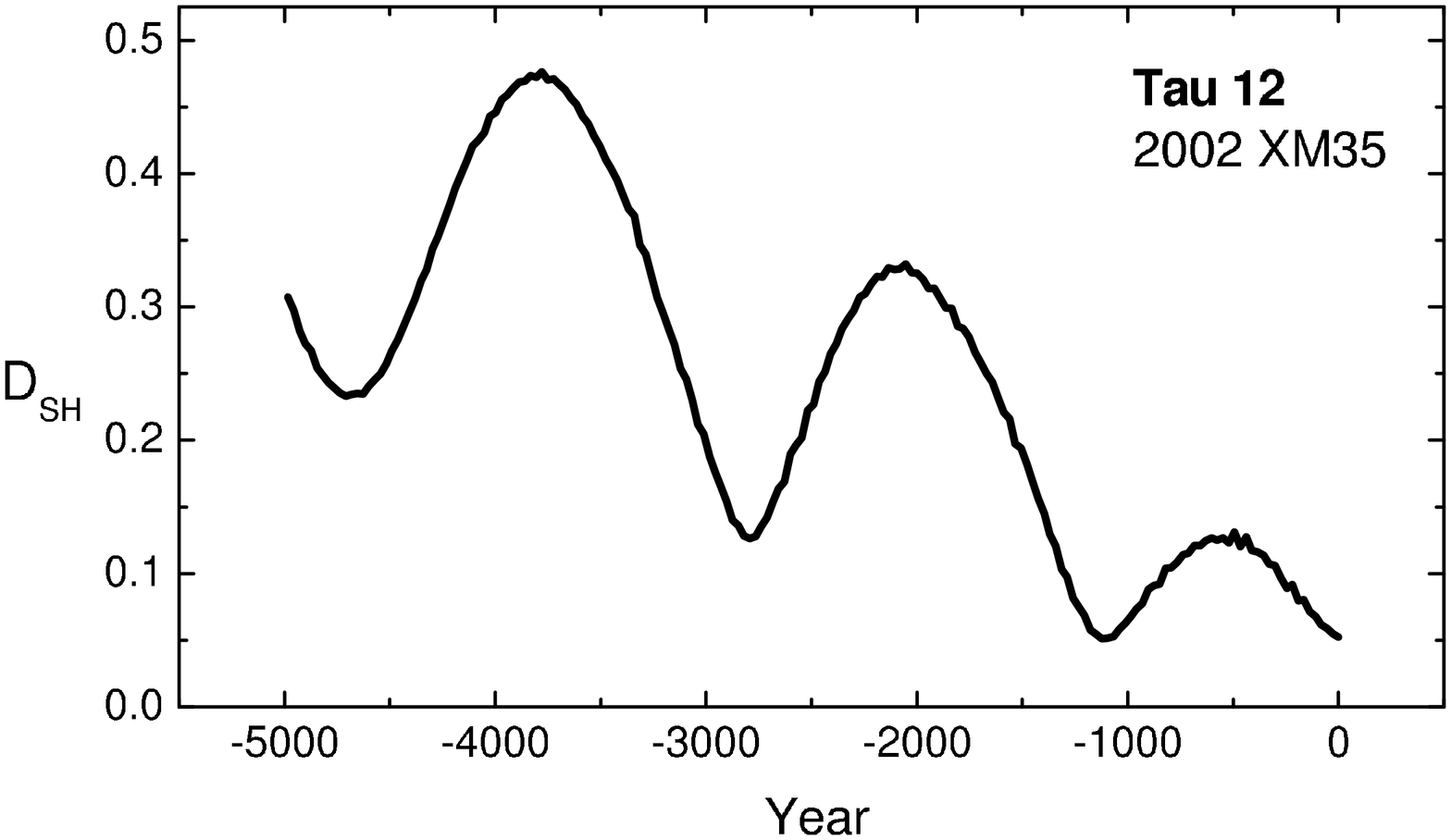}
            \hspace{0.3cm}
            \includegraphics[width=3.3cm,angle=-90]{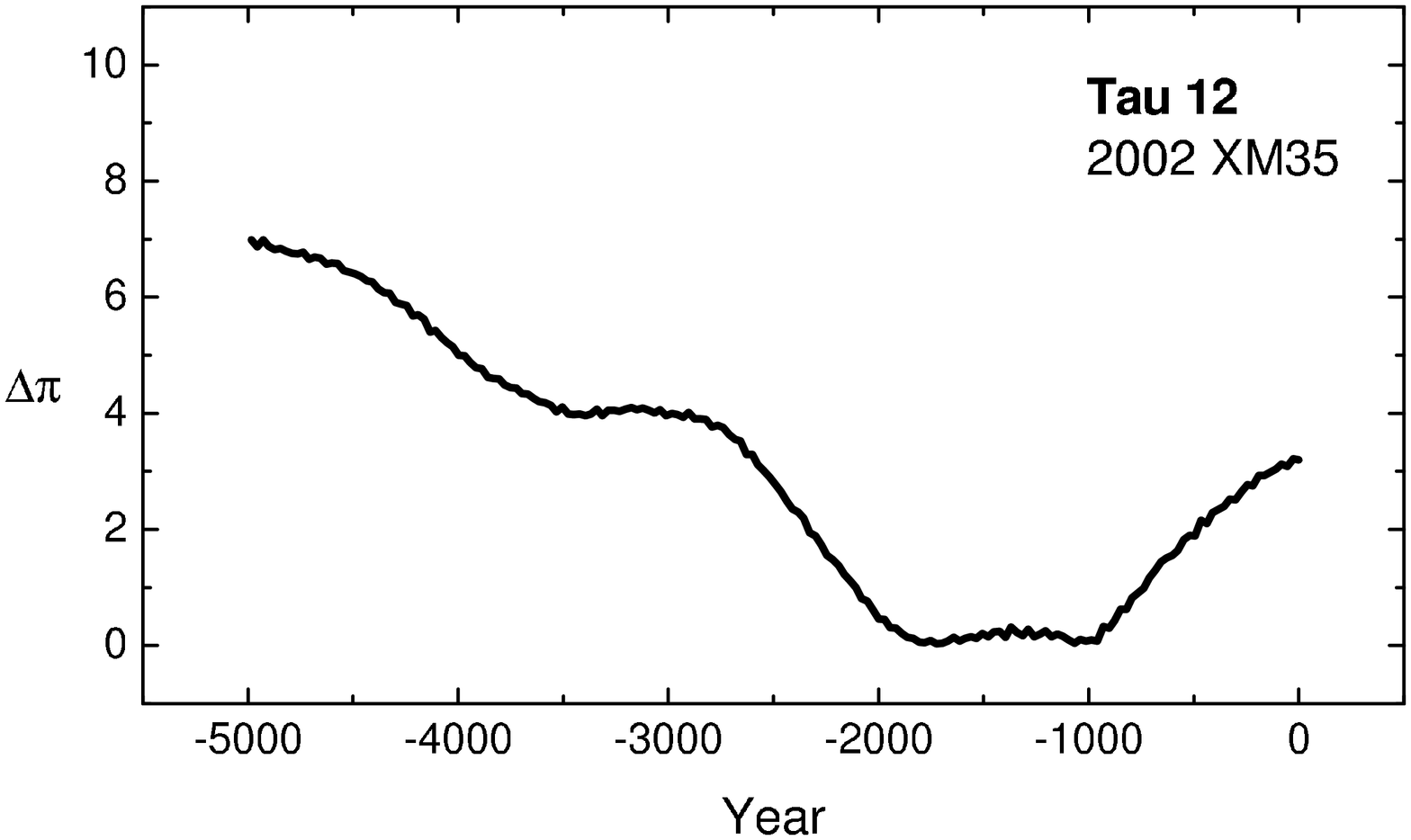}
           }
\caption{Orbital evolution and $D_{SH}$ of TC filament 12 and 2002 XM35.}
\end{figure}
\clearpage

%Fig. Encke
\begin {figure}
\centerline{\includegraphics[width=3.3cm,angle=-90]{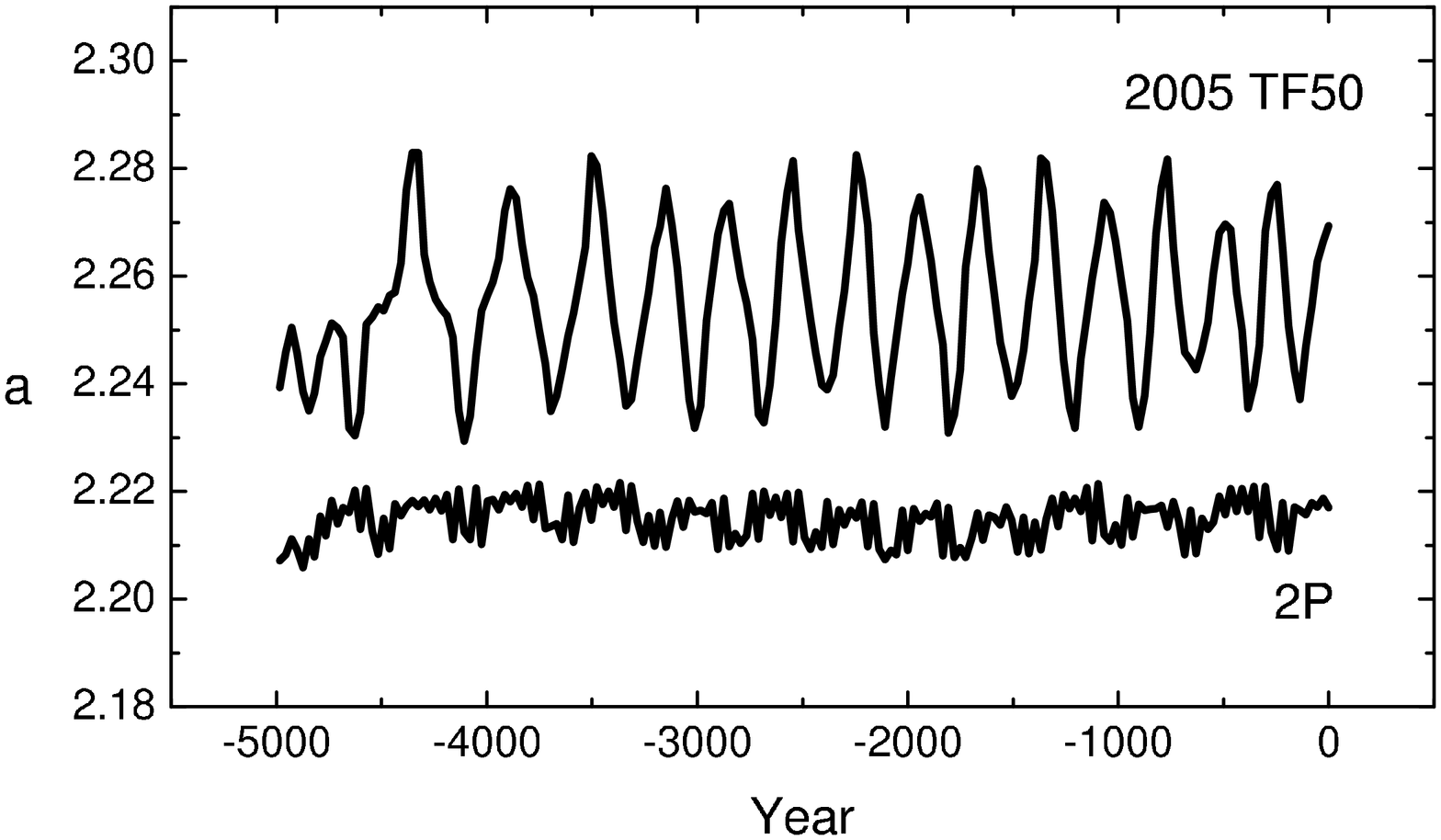}
           \hspace{0.3cm}
            \includegraphics[width=3.3cm,angle=-90]{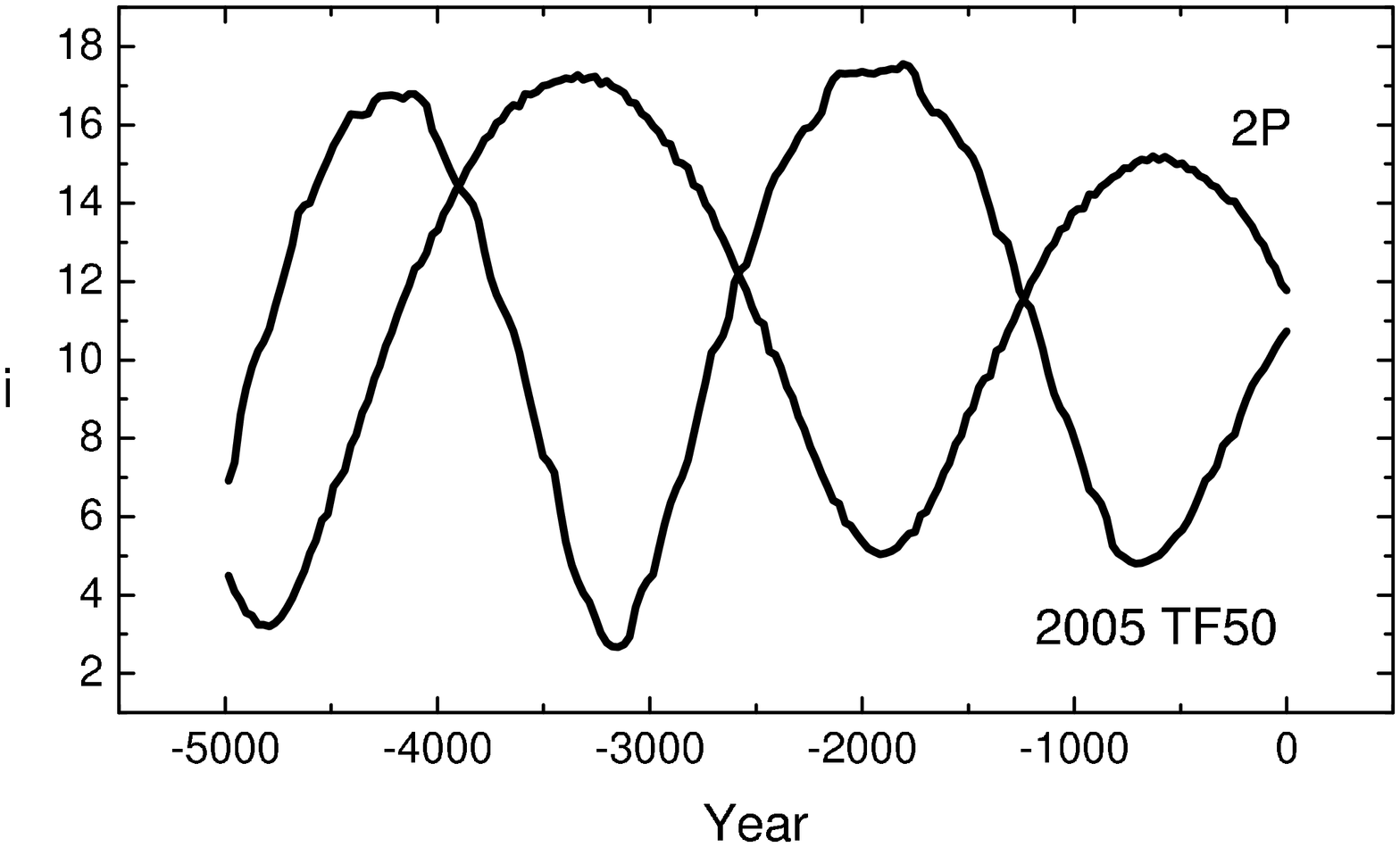}
           }
\vspace{0.7cm}
\centerline{\includegraphics[width=3.3cm,angle=-90]{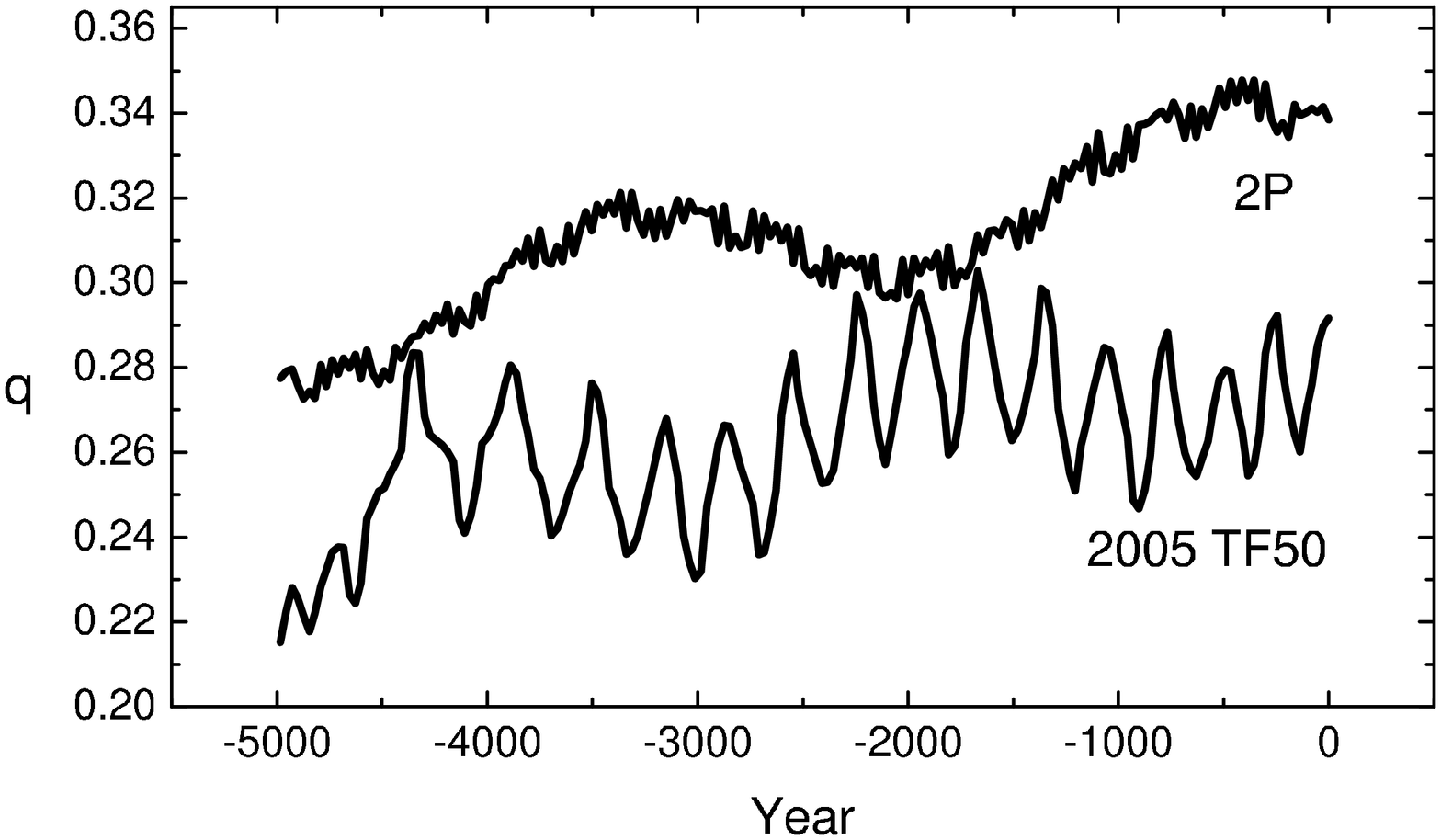}
            \hspace{0.3cm}
            \includegraphics[width=3.3cm,angle=-90]{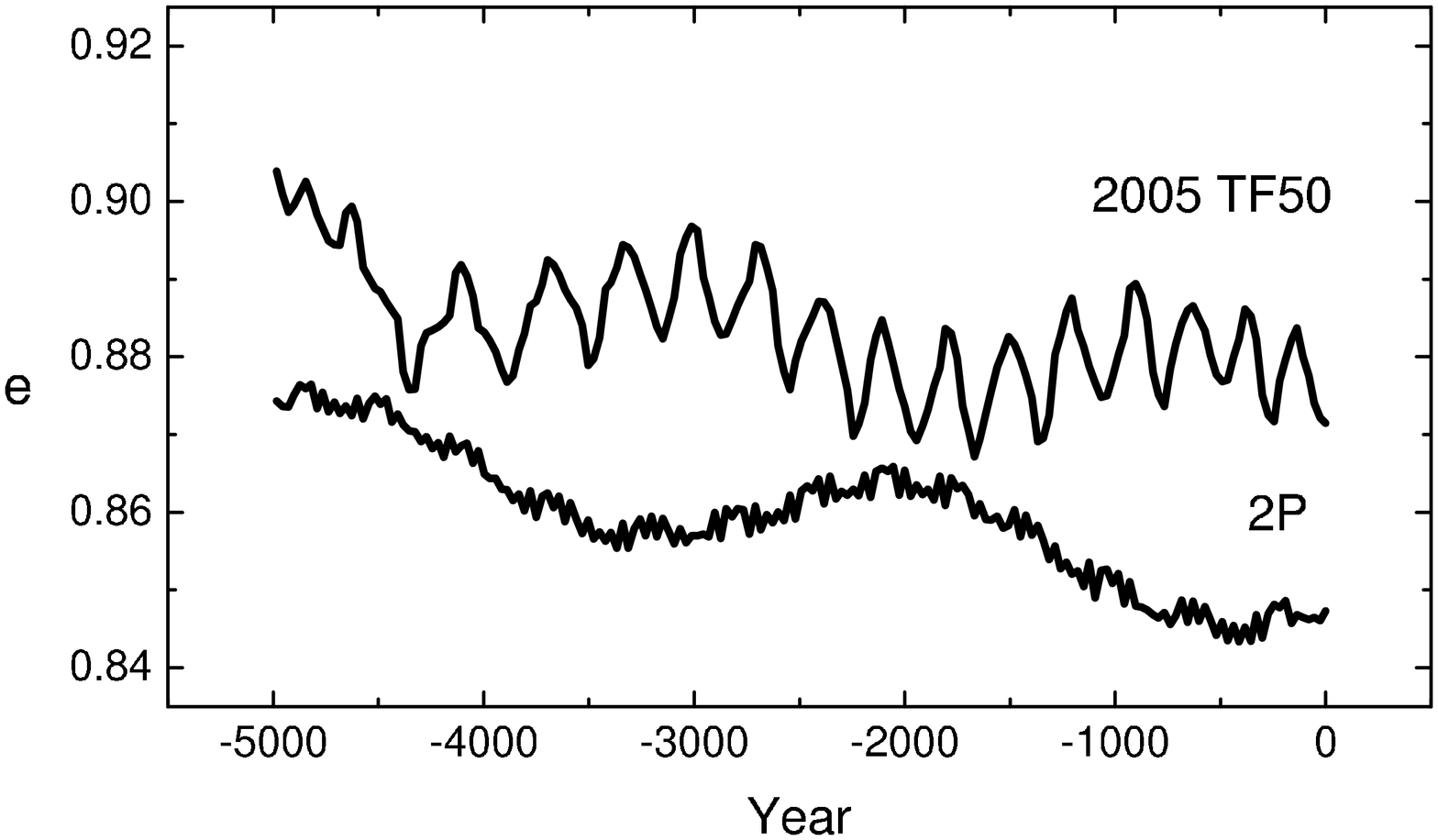}
           }
\vspace{0.7cm}
\centerline{\includegraphics[width=3.3cm,angle=-90]{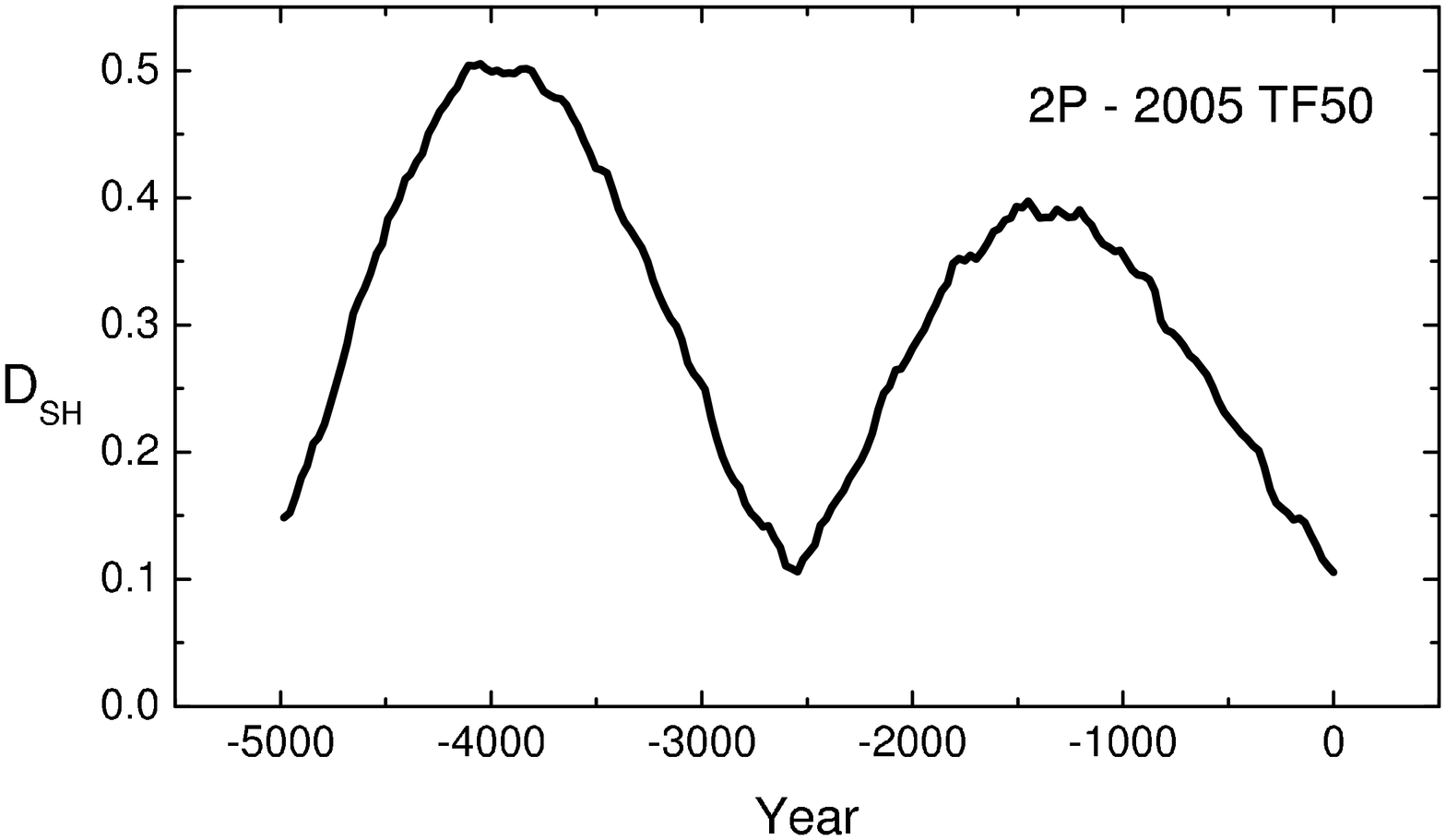}
            \hspace{0.3cm}
            \includegraphics[width=3.3cm,angle=-90]{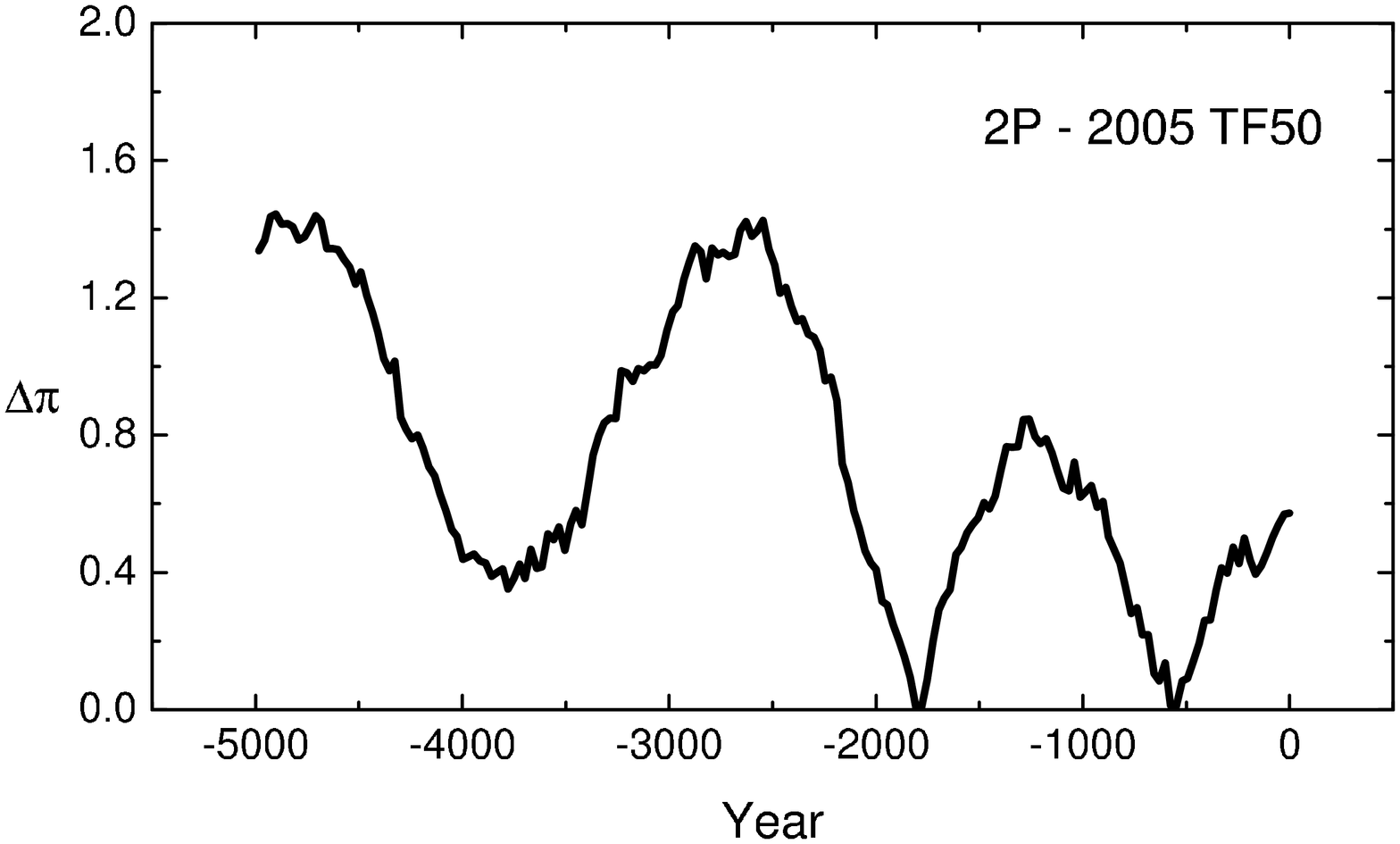}
           }
\vspace{0.7cm}
\centerline{\includegraphics[width=3.3cm,angle=-90]{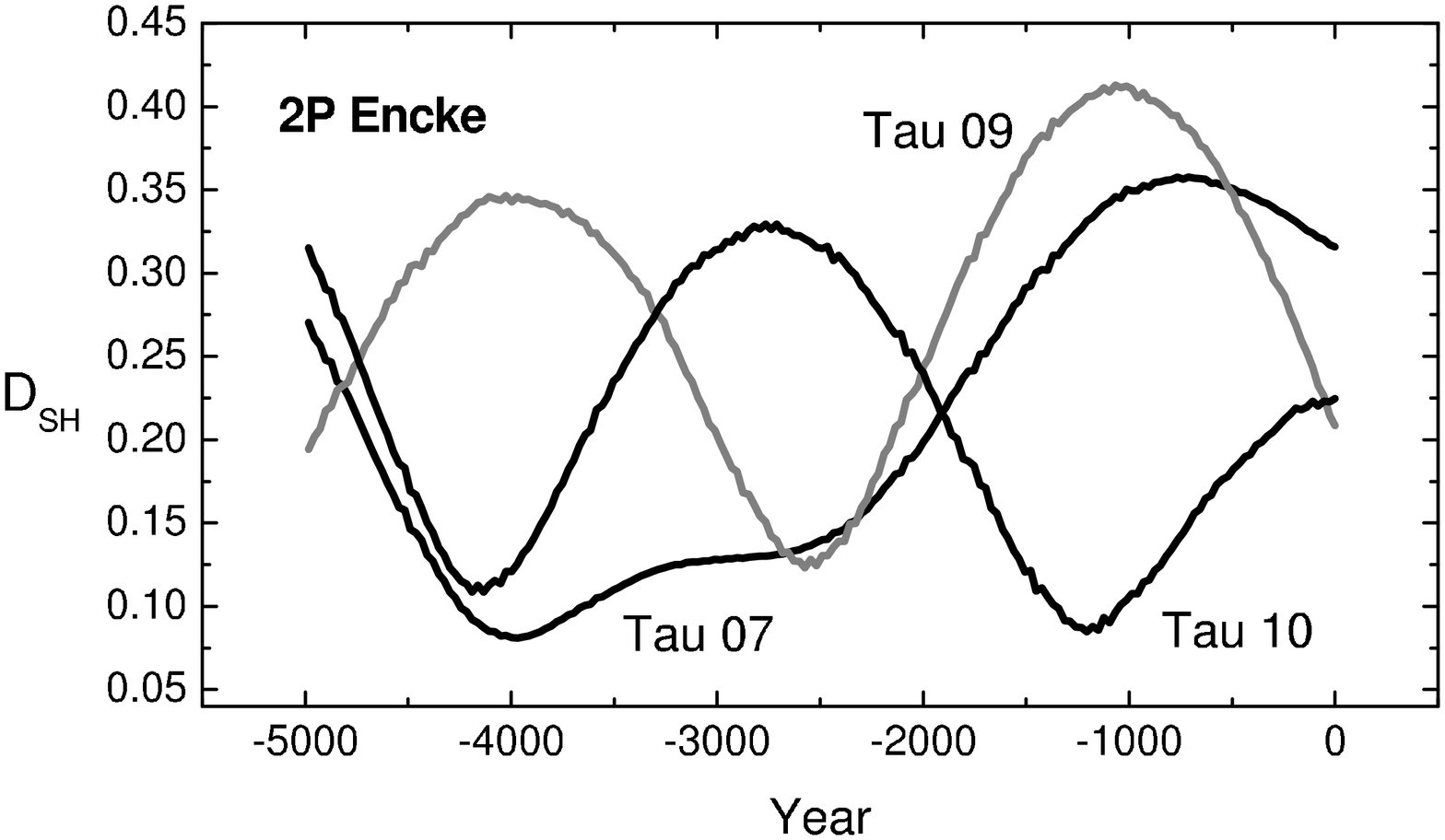}
            \hspace{0.3cm}
            \includegraphics[width=3.3cm,angle=-90]{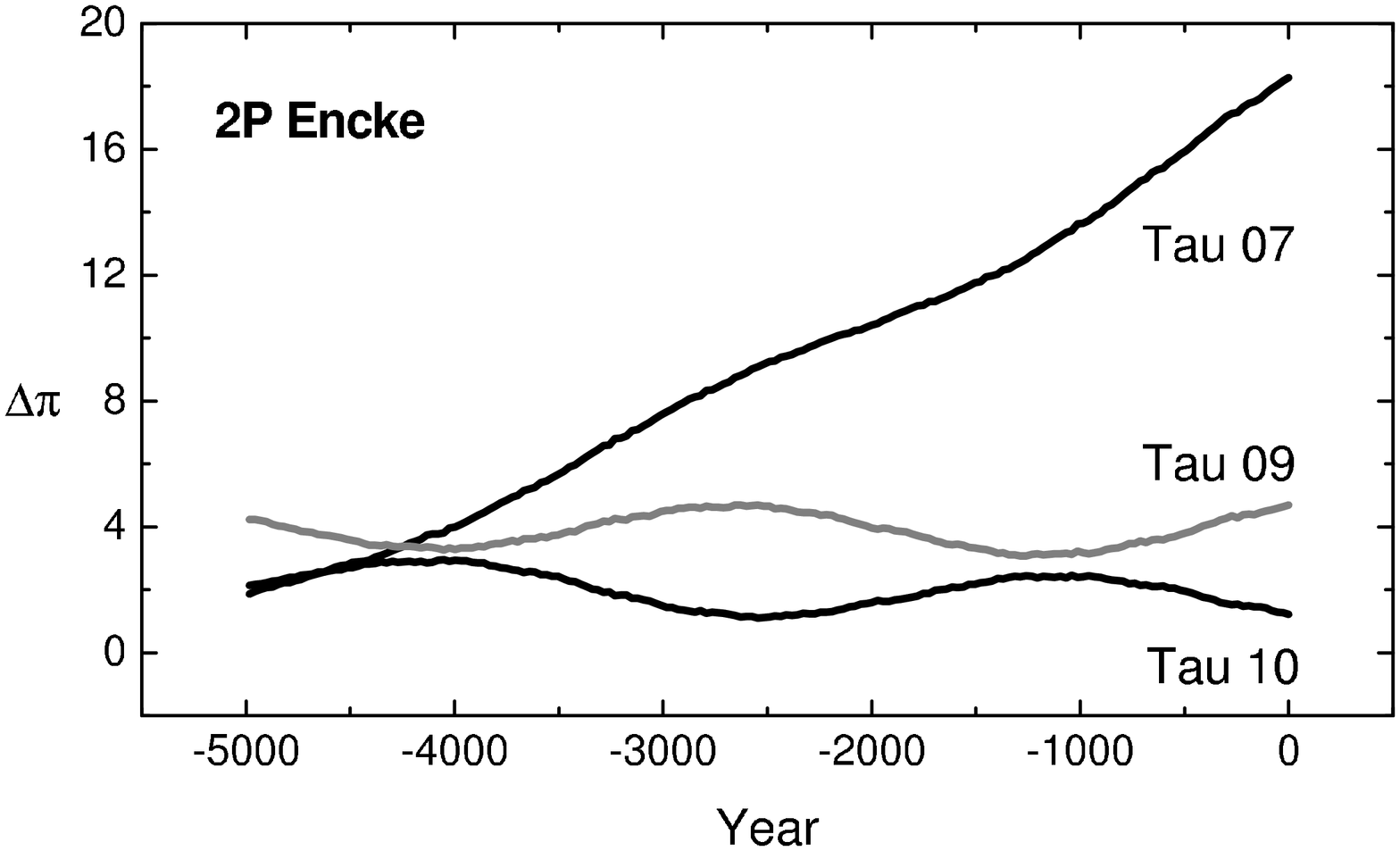}
           }
\caption{Orbital evolution and $D_{SH}$ of 2P/Encke and 2005 TF50 (upper six
plots) and evolution in $D_{SH}$ and $\Delta\pi$ between 2P/Encke and Taurid
filaments 7, 9 and 10 (lower two plots).}
\end{figure}

The most important of the elements when considering orbital
evolution are the semimajor axis $a$ and
aphelion distance $Q$ which characterize the distance of the
filament from the Jupiter's orbit. These influence the amplitudes
and rates of changes of the elements $q$, $e$ and $i$ during the
integration. The first two groups (I and II) have Tisserand
invariant close to P/Encke ($T = 3.03$) while the groups III and
IV are with their $a$, $Q$ and $T$ on typically asteroidal orbits.

Filaments 6 and 15 (group II) are close to 4\,:\,1 mean motion resonance with
Jupiter and this causes a large dispersion of the modeled particles in the
evolution of both streams. Filament 14 has a very high aphelion distance (4.9
AU), the particles approach Jupiter and many of them are disturbed.

The best defined association is for filament 4 and 2003\,QC10 with a very low $D$
for the whole period of integration and the association is on a typical
asteroidal orbit.  The next one is the association between filament 10 and
2004\,TG10 which has $D\le0.1$ for the last 2200 years and their lines of
apsides are very close for the whole period of integration.

Whipple's suggestion that comet Encke is one fragment of a giant comet which
disintegrated in the distant past gained support when in October 2005 asteroid 2005
TF50 was discovered  (Spahr, 2005). The asteroid is moving close to Encke
(Tab.\,3). Though the orbit of 2005\,TF50 is still rather uncertain,  the
orbital evolution of comet Encke and 2005\,TF50 (not considering
non-gravitational effects) is calculated and shown in Fig.\,7 (upper six
plots). 2005\,TF50 is close to 7\,:\,2 resonance with Jupiter.

Following the orbital evolution of 2P/Encke and  Taurid filaments, the closest
similarity in evolution between the comet and filaments is for filaments 7, 9
and 10, which contain 28, 56 and 38 meteors (Tab.\,1) the richest filaments
of the complex and correspond to the early S Tau (7), main S Tau (9) and main N
Tau (10) branches of the complex. All the three filaments have the Tisserand
invariant value close to that of Encke. The last two plots of Fig.\,7 show comparison
of Encke and the filaments as evolution in the differences of $D_{SH}$ and
$\Delta\pi$. The orientation of the lines of apsides of filaments 9 and 10 with
respect to Encke remains practically the same over the whole period of integration
and all three filaments had the same orientation about 4000 -- 4500 years ago.
At the same time small $D_{SH}$ values for filament 7 and 10 in this period may
suggest a possible origin of both filaments from Encke at this time. Another
enrichment of filament 10 from comet Encke may have occured about 1200 years
ago. Filament 9 could have originated from Encke about 2500 years ago (smallest
$D_{SH}$) but more probably earlier, more than 5000 years ago.

Summarizing, the following conclusions can be drawn:

Applying a strict value of the Southworth-Hawkins D-criterion
($D\le0.1$) to photographic orbits of the Meteor Data Center
catalogue - version 2003, 23 TC filaments containing from 2 to 56
members (210 meteors) were identified.

15 Taurid meteor complex filaments contain more than three meteors each and
these were searched for potential parents. The complex forms  a very broad
stream active for almost four months (Fig.\,2), 120$\degr$ in solar longitude.
The radiant area of the complex reduced to the common solar longitude  of
$220\degr$ is $25\degr\times15\degr$.

From the data set of 3380 NEO known until June 2005, 91 objects
were found to move close to the TC (for $D\leq0.30$).

Following the orbital evolution backwards over 5000 years, possible
associations between 7 TC filaments and 9 NEOs were found and the most probable
are for filaments (04): S Psc(b) -- 2003\,QC10, (10): N Tau(a) -- 2004\,TG10,
(11): $o$ Ori -- 2003\,UL3 and (12): N Tau(b) -- 2002\,XM35.

All the NEO in the TC are small objects with diameters less than 2 km
and may be collisional fragments, dormant cometary nuclei, asteroidal
fragments or larger meteoroids.

The most close associations between 2P/Encke and Taurid filaments were found
for filaments 7 (the earlier branch of S Tau), 9 (the main branch of S Tau) and
10 (the main branch of N Tau). Filaments 7 and 10 originated from the comet
about 4000 -- 4500 years ago and filament 10 was enriched about 1200 years ago,
filament 9 originated from comet probably still before 5000 years ago.

\acknowledgements
The authors acknowledge valuable comments from the referee and
support from the Slovak Grant Agency, Grant No. 1/0204/03.

\end{document}